\documentclass[%
 reprint,
noshowpacs,%preprintnumbers,
 amsmath,amssymb,
 aps,
prb,
floatfix
]{revtex4-1}

\usepackage{graphicx}% Include figure files
\usepackage{epstopdf}
\usepackage{dcolumn}% Align table columns on decimal point
\usepackage{bm}% bold math
\usepackage{makecell}
\usepackage{xcolor}
\begin{document}

\title
{
  Theory of electric field breakdown nucleation due to mobile dislocations
}
\author{Eliyahu Zvi Engelberg}
\author{Ayelet \surname{Badichi Yashar}}
\author{Yinon Ashkenazy}
\author{Michael Assaf}
\author{Inna Popov}
\affiliation{
  Racah Institute of Physics and
  the Center for Nanoscience and Nanotechnology,
  The Hebrew University of Jerusalem,
  Jerusalem 9190401, Israel
}
\date{\today}

\begin{abstract}
  A model is described,
  in which electrical breakdown in high-voltage systems is caused by stochastic
  fluctuations of the mobile dislocation population in the cathode.
  In this model, the mobile dislocation density normally fluctuates,
  with a finite probability to undergo a critical transition due to the effects
  of the external field.
  It is suggested that once such a transition occurs,
  the mobile dislocation density will increase deterministically,
  leading to electrical breakdown.
  Model parametrization is achieved via microscopic analysis
  of OFHC Cu cathode samples from the CERN CLIC project,
  allowing the creation and depletion rates of mobile dislocations to be
  estimated as a function of the initial physical condition of the material
  and the applied electric field.
  We find analytical expressions for the mean breakdown time and quasistationary
  probability distribution of the mobile dislocation density,
  and verify these results by using a Gillespie algorithm.
  A least-squares algorithm is used to fit these results with
  available experimental data of the dependence of the breakdown rate on the
  applied strength of the electric field and on temperature.
  The effects of the variation of some of the assumptions of the physical model
  are considered,
  and a number of additional experiments to validate the model are proposed,
  which include examining the effects of the temperature and pulse length,
  as well as of a time-dependent electric field,
  on the breakdown rate.
  Finally, applications of the model are discussed,
  including the usage of the quasistatic probability distribution
  to predict breakdowns,
  and applying the predictions of the model to
  improve the conditioning process of the cathode material.
\end{abstract}

\maketitle

\section{Introduction} \label{sec_introduction}

The process of plastic deformation in metals is known to be controlled by
dislocation dynamics.\cite{taylor34, hirth85}
Due to the stochastic nature of these dynamics,
plastic deformation can occur even below the yield point of the metal.
For example,
aging can be observed in metals subjected to cyclic low stresses,
due to collective stochastic motion of dislocations.
These may lead to strain localization and formation of structures
known as persistent slip bands.
In particular,
the formation of surface features occurs at the surface-slip band
intersection.\cite{blum09, man09, levitin09, goto08, laurent11}

Even at stresses close to the yield point,
a complete analysis of the dislocation dynamics
must take into account the stochastic nature of mobile dislocation
nucleation and depletion.\cite{Meyers_2002}
For instance,
it was shown experimentally and through simulation that the
compression of micropillars,
which can be formed as single crystals with a low dislocation density,
consists of a series of discrete slip events,
in which applied stress unpins sessile dislocations
and enables them to move to the surface of the crystal.\cite{Dimiduk_2009}
The probability distribution of such events was measured\cite{dimiduk06}
and shown to match simulations,\cite{csikor07}
and mean field theories were developed which were able
to reproduce the stress-strain behavior
of the micropillars.\cite{Dahmen_2012, Nix_Lee_2011, ryu13, nix_gao_2015}

In a previous study\cite{engelberg18} we proposed that stochastic fluctuations
of the mobile dislocation density $\rho$ control a critical process in
metallic surfaces subjected to an extreme electric field.
This critical process leads to plasma formation between electrodes in vacuum,
and to subsequent arcing of current between the electrodes,
serving as a major failure mechanism in numerous
applications.\cite{boxman96, slade08, gai14, keidar}
Specifically, arcing between electrodes, known as breakdown,
limits the design of linear accelerators,
and as such is a focal topic of the prospect study for a future compact linear
accelerator (CLIC) in CERN.\cite{clic_stage}
CLIC is planned to operate at low breakdown rates (BDRs)
with electric fields of 100 MV/m and stronger applied between
OFHC Cu electrodes.
Since a large amount of experimental results and physical samples
from the CLIC project are available for analysis,
this manuscript focuses on theoretical estimates for OFHC Cu,
the parameters of which are directly derived using samples
from the CLIC project.
Results from CLIC consist of data collected both from setups
where short radio frequency (RF) electromagnetic pulses are
applied,\cite{grudiev09}
and from setups in which the electric field is constant
(DC).\cite{descoeudres09}

\begin{figure}
  \centering
  \includegraphics[width=8.5cm]{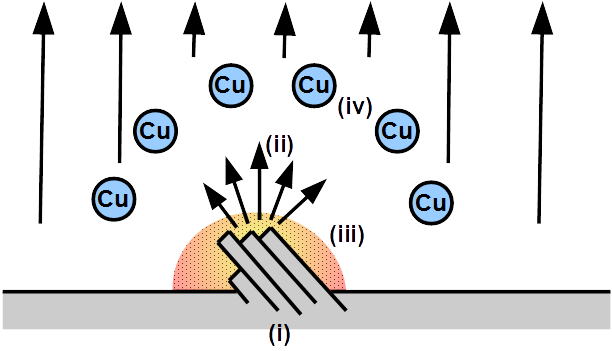}
  \caption
  {
    \label{fig_process}
    Schematic description of the stages leading to arcing:
    (i) Extreme fields generate local stresses which, in turn,
    lead to dislocation activity,
    causing mobile dislocations to glide to the surface of the metal
    and modify it.
    (ii) The electric field and current on the surface are enhanced due to the
    surface modifications.
    (iii) The enhanced current causes heating.
    (iv) Due to the heating,
    atoms are released from the cathode and plasma is formed,
    allowing current to arc between the electrodes.
  }
\end{figure}

The process of arc nucleation under extreme electric fields is understood to be
initiated by the glide of mobile dislocations to the surface of the metal,
due to local stresses generated by the fields.\cite{calatroni10}
The mobile dislocations arriving at the surface modify it,
thus enhancing the electric field and the current on the
surface.\cite{pohjonen11}
The enhanced current then causes heating,
which causes atoms to be released from the cathode and to form
a plasma,\cite{anders08, anders14}
allowing current to arc between the electrodes.\cite{boxman96}
This process is described schematically in Fig.~\ref{fig_process}.

Previous attempts to explain breakdown nucleation
were centered around the formation
of distinct protrusions leading to electric field enhancement,
evidenced by increased dark currents.\cite{zadin14, vigonski15}
The enhanced electric field can lead to heating,
  due to the current and field emission effects after a significant
  surface protrusion appears.\cite{kyritsakis18}

However,
the formation process of such protrusions in a metal subject to an electric
field has not yet been adequately
described theoretically or observed
experimentally.\cite{pohjonen11, zadin14, gai14, vigonski15}

Our model,
based on mobile dislocation density fluctuations (MDDF),\cite{engelberg18}
complements these previous models by proposing that surface features
appear as a result of a critical increase
in the mobile dislocation density $\rho$.
According to this model, prior to breakdown,
the mobile dislocation density is in a long-lived metastable state,
fluctuating around a deterministically stable value $\rho_*$.
When the population experiences a large enough fluctuation to carry the
mobile dislocation density beyond a critical value $\rho_c$,
a critical transition occurs,
leading to a deterministic increase in the mobile dislocation density,
which can lead to a localized increase in field emission,
  due to plastic evolution of the surface.
Therefore,
the MDDF model describes the process up to the formation of surface deformations,
while the subsequent processes of breakdown can be treated by the models
previously mentioned.\cite{pohjonen11, zadin14, gai14, vigonski15}
This post-nucleation evolution is not discussed here,
  and may, as well, not be deterministic.
  Indeed,
  there are initial indications from microscopy and current measurements
  suggesting the existence of sub-breakdown events,
  which may be a result of critical transitions which did not develop into a
  full-blown breakdown.\cite{werner04}

In this study we extend the MDDF model\cite{engelberg18}
by including insights from experimental observations
pertaining to its physical characteristics and parameters,
and discussing their implications for the model.
In addition,
we present predictions of the model which are relevant
for applications in which electric field breakdown is significant.

The manuscript is organized as follows:
In Section \ref{sec_mean} we present the physical basis of the model,
consisting of deterministic rate equations describing the creation and depletion
of mobile dislocations in a metal subjected to an electric field.
Then, in Section \ref{sec_stochastic},
we describe the problem of finding the BDR in terms of a birth-death master
equation\cite{gardiner04} for the mobile dislocation population,
thereby transforming the problem of calculating the BDR to that
of finding the first passage time of a biased random walker.
Results of the model are compared to experimental measurements of BDRs
in OFHC Cu in Section \ref{sec_model},
providing estimates of observables
such as the activation energy and volume for mobile dislocation nucleation.
In Section \ref{sec_sensitivity} we examine variations
of the physical assumptions of the model,
and demonstrate the robustness of the resulting BDR dependence
on the electric field.
In Section \ref{sec_proposed} we propose specific experiments,
which can serve to validate the predictions of the model.
Finally, in Section \ref{sec_discussion},
possibilities for reducing the BDR in real-life applications are discussed.

\section{Mean-field model} \label{sec_mean}

\subsection{Kinetic equations}

Under externally applied stress,
dislocations will glide along slip planes.\cite{taylor34}
The resulting mobile dislocation density $\rho$ is expressed as the total length
of dislocations in one slip plane,
and therefore measured in units of $\text{nm}^{-1}$.
Mobile dislocations can be blocked by various obstacles,
including interactions with other dislocations.
Once rendered sessile,
dislocations can be released by processes such as cross slip.\cite{hirth82}
Thus, barriers serve both as sources and sinks of mobile dislocations.

\begin{figure}
  \centering
  \includegraphics[width=8.5cm]{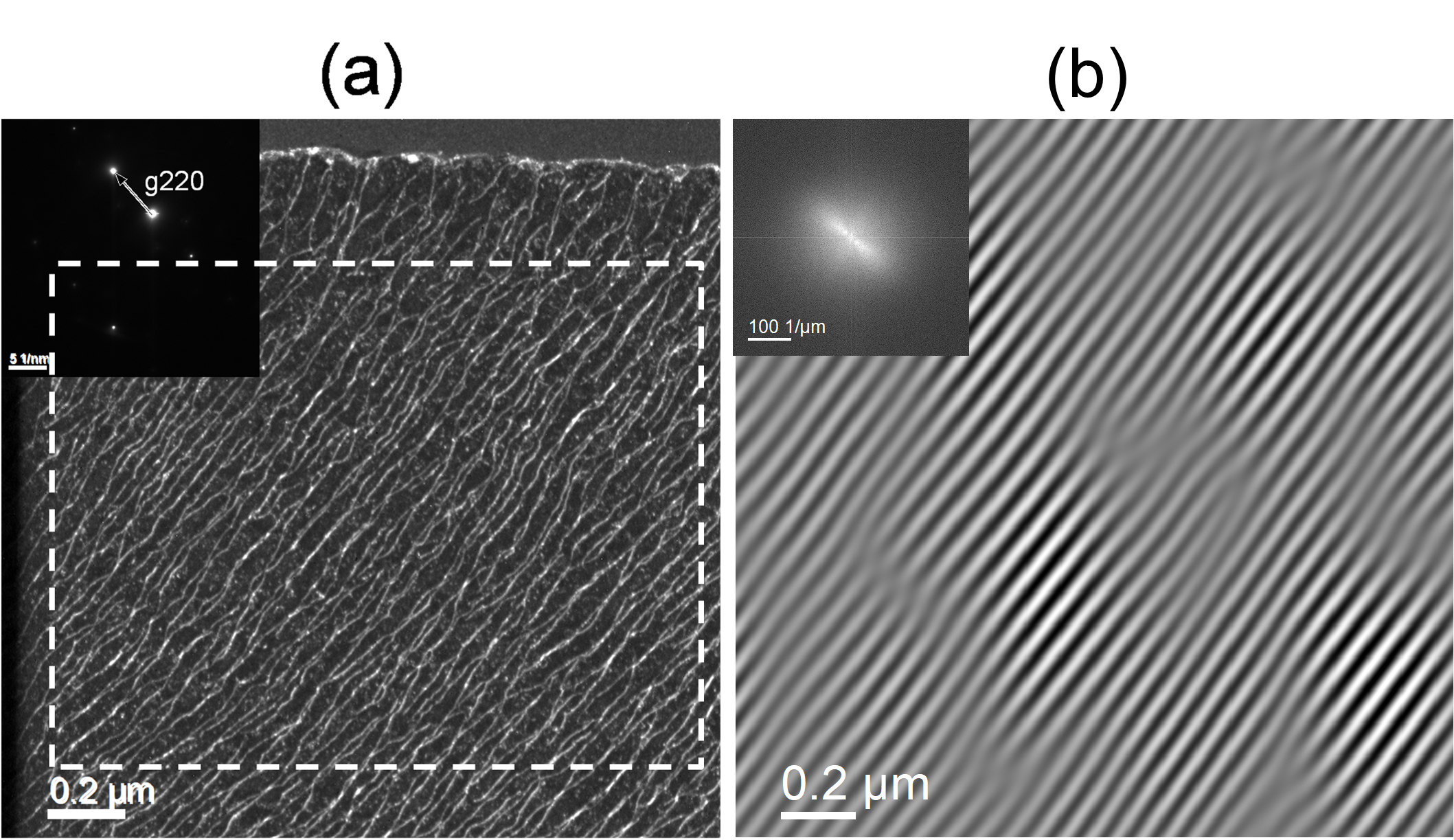}
  \caption
  {
    \label{fig_sessile_dislocations}
    (a) A dark-field TEM image of a soft OFHC Cu sample
    ($\sim 100$ nm thick lamella) under two-beam g:220 diffraction conditions,
    displaying a typical ladder-like dislocation structure,
    with the corresponding diffraction pattern (upper left corner).
    (b) Fourier filtered region enclosed by the dashed rectangle in (a),
    spatial frequencies farther away than $\sim 10\,\mu\text{m}^{-1}$
    from the peak spatial frequency filtered out.
    The FFT of the region is shown in the upper left corner.
    The peak spatial frequency corresponds to a transverse distance of 62 nm
    between dislocation lines.
  }
\end{figure}

Figure \ref{fig_sessile_dislocations} is a dark-field transmission electron
microscopy (TEM) image of a soft OFHC Cu cathode sample from CLIC,
under two-beam g:220 diffraction conditions.
Dislocation lines are seen to be aligned in a typical ladder-like
structure\cite{mughrabi92} in one active slip plane,
which is parallel to the image plane,
and separated from each other by a transverse distance of $\sim 62$ nm.
The density of barriers $c$ can be estimated from the observed distance between
dislocation intersections,
which is seen in the figure to be of the order of $\sim 1\,\mu$m.

To find the creation and depletion rates of mobile dislocations,
we consider a zero-dimensional mean-field model,
in which the average value of the mobile dislocation density in one slip plane
is calculated.
Thus,
the effects of variation of the mobile and sessile dislocation density within
the slip plane,
which would lead to spatial variation of the rates of creation and depletion,
are neglected.
Also,
the variation of the dislocation density and cross interactions
among slip planes are not taken into consideration.
Cross-interaction effects due to localized surface evolution are
  not considered,
  either,
  as these are expected to evolve only following the nucleation of a breakdown
  event.

When an electromagnetic field is applied,
the rate at which new mobile dislocations are created is, therefore,
determined by the longitudinal density of barriers $c$ within the slip plane,
and is proportional to the rate of creation of mobile dislocations
at each barrier.
Since the creation of mobile dislocations is thermally activated,
the creation rate should also be proportional to a temperature-dependent
factor $\exp[-(E_a - \Omega\sigma) / (k_B T)]$,
divided by the average creation time of each mobile dislocation.
Here $E_a$ and $\Omega$ are the activation energy and volume, respectively,
of a mobile dislocation nucleation source,
whose values we estimate in Section \ref{sec_model},
while $\sigma$ is the stress within the slip plane.

The average creation time $t$ is calculated by considering
a typical Frank-Read type source.\cite{frank50}
In such a case $t = L/v$,
with $L$ the length of the source,
and $v$ the velocity of the mobile dislocation.
The threshold stress needed to activate such a source is
$\sigma_\text{th} = 2Gb/L$,
where $G = 48$ GPa is the shear modulus,
and $b = 0.25$ nm is the Burgers vector.\cite{weertman64}
If the amount of sources decreases rapidly as a function of length,
then the dislocation sources can be described using a single
$L \approx 2Gb/\sigma$.
For stresses ranging from 0.2 MPa,\cite{greenman67, nadgornyi88}
up to 400 MPa,\cite{hirth82, ashkenazy03}
the dislocation velocity in Cu is approximately a linear function of $\sigma$,
$v = 50 C_t \sigma/G$,
where $C_t = 2.31\times 10^3$ m/s
is the propagation velocity of sound in Cu.\cite{hirth82}
Therefore the average creation time satisfies, $t = G^2b / (25 C_t \sigma^2)$,
giving us a total creation rate
\begin{equation}
  \dot{\rho}^+ =
  \frac{25\kappa C_t c}{G^2 b} \sigma^2
  \exp\left(-\frac{E_a - \Omega\sigma}{k_B T}\right), \label{eq_drhop}
\end{equation}
where $\kappa$ is a kinetic factor which depends on the
activation entropy of the sources,\cite{ryu11}
evaluated in Section \ref{sec_model}.

To estimate the depletion rate of mobile dislocations,
we consider dislocation arrest at barriers and sufaces.
Assuming that the latter mechanism is considerably slower than the former,
we can approximate the depletion rate as $\dot{\rho}^- = \xi c \rho v$.
Here $\xi$ is a dimensionless proportionality factor,
representing trap efficiency.
For simplicity, we assign it a value of 1.
Substituting once again for the dislocation velocity $v$ we have
\begin{equation}
  \dot{\rho}^- = \frac{50\xi C_t c}{G} \sigma\rho. \label{eq_drhom}
\end{equation}

\subsection{In-plane stress}

The stress in a slip plane is composed of two terms:
the Maxwell stress due to the applied electormagnetic fields
\textbf{E} and \textbf{B},
and the internal stress caused by the dislocations themselves.
The Maxwell stress in each direction,
i.e., the force in each Cartesian direction per unit area of the surface
acting on the particles and fields inside the metal,
is given as $\sum_\beta T_{\alpha\beta}n_\beta$,
with
$T_{\alpha\beta} = \epsilon_0[E_\alpha E_\beta + c^2 B_\alpha B_\beta
  - \frac{1}{2}(E^2 + c^2 B^2)\delta_{\alpha\beta}]$,
where $E_i$ and $B_i$ are the Cartesian components
of the electric and magnetic field,
and $E$ and $B$ are their respective magnitudes.\cite{jackson99}
In the case of a static electric field (DC),
the stress inside the slip plane, close to the surface,
can be estimated to have a uniform value of $\epsilon_0 (\beta E)^2 / 2$
in the direction perpendicular to the surface.\cite{pohjonen11}
Here,
the dimensionless parameter $\beta$ represents
the ratio of the average stress inside the slip plane to the stress
on the surface.
$\beta$ is expected to depend on both surface geometry
and the electric field distribution.
Specifically, one may expect $\beta$ to vary with $\rho$,
since it relates to plastic deformation of the
surface.\cite{chatterton66, wang04, descoeudres09}
However,
due to the low variation range of $\rho$ prior to breakdown,
we consider $\beta$ to be constant per cathode geometry
(see Section \ref{sec_model}).
This is consistent with the fact that no surface evolution
was microscopically observed in pre-breakdown samples,
as described in Section \ref{sec_introduction}.

In the case of an alternating electric field (RF),
the in-plane stress includes magnetic field terms,
and components of the electric field parallel to the surface,
in addition to the contribution of the perpendicular electric field.
Above a frequency of 1 GHz, and at subyield stresses,
the effect of the fields on the mobility of dislocations
can be estimated using the average of the fields over time.
The additional components of the stress are then linearly proportional
to the perpendicular field,
so that their contribution can be incorporated into the value of $\beta$.

Note that,
  due the nature of the Maxwell stress tensor $T_{\alpha\beta}$,
  stress will be induced in a metal even when it is subject only to a magnetic
  field.
  In such a scenario,\cite{laurent11} then,
  breakdown nucleation should be ultimately attributed
  to the applied magnetic field,
  since the effects of temperature alone, within experimental ranges,
  cannot account for breakdown on an initially smooth
  surface.\cite{kyritsakis18}

The second term of the stress,
due to the internal stress caused by the dislocations,
is proportional to $Gb/d$,
where $d$ is the average distance
between dislocations.\cite{taylor34, progress80}
In the experimental setups examined in Section \ref{sec_model}
a pulsed electric field is applied,
and the BDR is constant over time.
Since there is no memory effect,
we assume a constant sessile dislocation population
whose contribution to the total stress from all slip planes saturates.
As a result, we take into consideration
only the stress caused by the mobile dislocations,
whose density varies over time.
In multi-slip-plane systems $d$ is proportional to $\rho^{-1/2}$,
with $\rho$ measured in units of
nm\textsuperscript{-2}.\cite{taylor34, progress80}
However, when considering only one slip plane as in our model,
we expect the relation to be $d \sim \rho^{-1}$,
with $\rho$ in units of nm\textsuperscript{-1}, as described above
(and also see Section \ref{sec_sensitivity}).
We therefore find that overall, the stress is
\begin{equation}
  \sigma = \epsilon_0 (\beta E)^2 / 2 + ZGb\rho, \label{eq_stress}
\end{equation}
where the dimensionless parameter $Z$, in the second term of the stress,
is a structural parameter linking the stress to the dislocation density.
For concreteness, we assign it a value of 1.

\subsection{Deterministic fixed points}

\begin{figure}
  \centering
  \includegraphics[width=8.5cm]{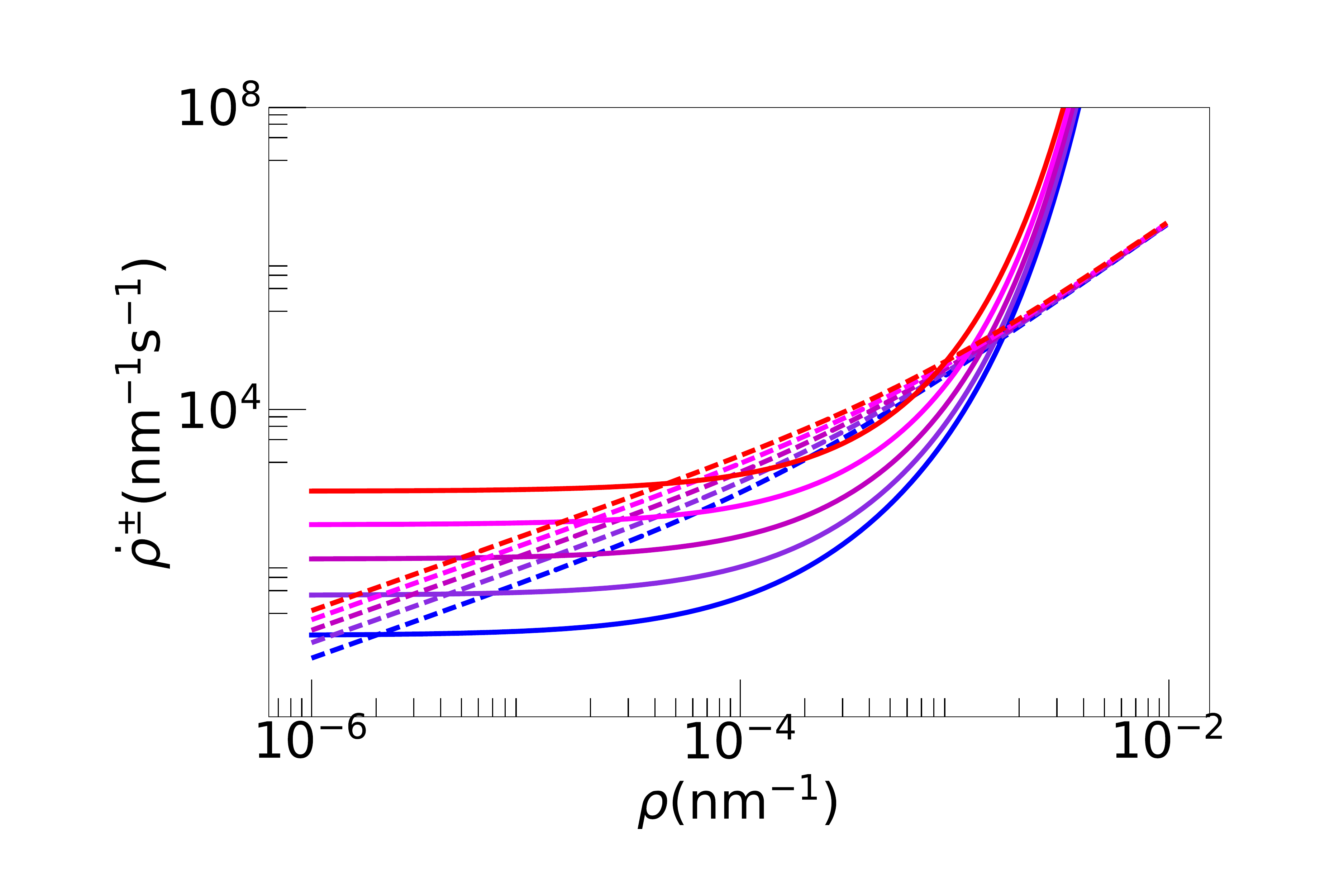}
  \caption
  {
    \label{fig_analytical_drhomp}
    $\dot{\rho}^+$ (solid lines) and $\dot{\rho}^-$ (dashed lines)
    for five electric fields (bottom to top):
    150, 190, 230, 270, and 310 MV/m.
  }
\end{figure}

Defining new constants of the form $\alpha = \Omega / (k_B T)$,
$A_1 = \epsilon_0 (\beta E)^2 / 2$, $a_2 = ZGb$,
$B_1 = 25\kappa C_t c\exp\left[-E_a / (k_B T)\right] / (G^2 b)$,
and $b_2 = 50\xi C_t c / G$,
we arrive at
\begin{equation}
  \label{eq_rho}
  \dot{\rho} = \dot{\rho}_+ - \dot{\rho}_-;\quad
  \dot{\rho}_+ = B_1\sigma^2 e^{\alpha\sigma},\quad
  \dot{\rho}_- = b_2\sigma\rho,
\end{equation}
with $\sigma = A_1 + a_2\rho$.
As can be seen,
$A_1$ is the only parameter that depends on the strength of the electric field.
The values of $E_a$ = 0.08$\pm$0.002 eV, $\Omega$ = 5.6$\pm$0.2 eV/GPa,
$\kappa$ = 0.32$\pm$0.02, and $\beta$ = 4.6$\pm$0.1,
found by the fitting procedure in Section \ref{sec_model},
give us the following values for the constants:
$A_1$ = 94 Pa\,(MV/m)\textsuperscript{-2}$E^2$, $a_2$ = 12 GPa\,nm,
$B_1$ = 0.15 Pa\textsuperscript{-2}\,m\textsuperscript{-1}\,s\textsuperscript{-1},
$b_2$ = 2.4 Pa\textsuperscript{-1}\,s\textsuperscript{-1},
and $\alpha$ = 220 GPa\textsuperscript{-1}.
Figure \ref{fig_analytical_drhomp} shows the values of $\dot{\rho}^+$
and $\dot{\rho}^-$ for these nominal values.
In the rest of this manuscript, unless stated otherwise,
the results presented are for these values.

The fixed points can be found in the following way:
For $\rho \ll A_1/a_2$,
we find a stable fixed point at $\rho_* = (B_1 A_1/b_2)e^{\alpha A_1}$,
while for $\rho \gg A_1/a_2$,
we find an unstable fixed point at
$\rho_c = (\alpha a_2)^{-1}\ln[b_2/(B_1 a_2)]$.
That is,
when $\rho_* < \rho < \rho_c$ we have $\dot{\rho}_- > \dot{\rho}_+$,
meaning that the mobile dislocation deterministically returns
to the stable attracting point $\rho_*$.
Whereas, when $\rho > \rho_c$,
we have $\dot{\rho}_+ > \dot{\rho}_-$,
meaning that the mobile dislocation density increases beyond $\rho_c$,
leading to eventual breakdown.

Note that as the electric field increases,
$\rho_*$ and $\rho_c$ approach each other,
and the assumption that $\rho_* \ll \rho_c$ becomes invalid.
The values of $\rho_*$ and $\rho_c$ coincide at a bifurcation point,
when the electric field is equal to the deterministic breakdown field $E_b$.
For $E > E_b$, we have $\dot{\rho}_+ > \dot{\rho}_-$ for every $\rho$.
Therefore, when a field greater than $E_b$ is applied,
the system does not possess a stable fixed point,
and it progresses directly to breakdown.

\subsection{Dislocation cells}

\begin{figure}
  \centering
  \includegraphics[width=8.5cm]{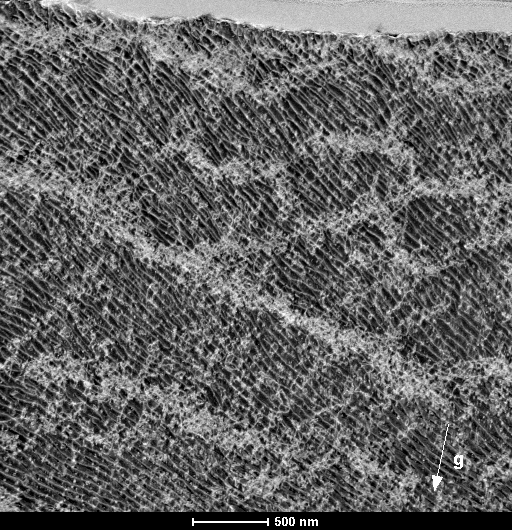}
  \caption
  {
    \label{fig_cells}
    A dark-field STEM image of a cross-sectional lamella from a fully
    conditioned (see main text) OFHC soft Cu electrode,
    under two-beam g:111 diffraction conditions,
    showing dislocation cells separated by dislocation walls.
    The cross section was taken from a region far (at least 50 $\mu$m away)
    from any breakdown site.
    The dislocation walls appear as curled thin bright lines,
    as opposed to long straight dislocation segments inside the cells
    which are organized in a ladder-like structure.
  }
\end{figure}

Under applied stress,
dislocations tend to become organized in a cellular structure,
where they are free to glide within each cell,
and the cells are separated by an accumulation of sessile
dislocations.\cite{amodeo88}
Figure \ref{fig_cells} shows a dark-field scanning transmission electron
microscopy (STEM) image of a cross-section taken from a soft OFHC Cu electrode
which was fully conditioned,
meaning that it was exposed to a pulsed electromagnetic field of increasing
intensity and pulse length,
so that its BDR reached a saturation value.
The borders of the cells appear as curled white lines,
where each cell is smaller than 10 $\mu$m.
The cross-section was taken from a region which is at least 50 $\mu$m away
from any breakdown site,
demonstrating that the formation of cells is a phenomenon caused by the stress
generated by the electric field,
and not by the breakdowns themselves.

Assuming the dislocation population evolves independently in each cell,
the addition or removal of a single mobile dislocation will
modify the mobile dislocation
density by approximately $\Delta\rho = 0.1\,\mu\text{m}^{-1}$.
Since breakdown is a surface phenomenon,
we propose that it is driven by the mobile dislocation population behavior
in the cells adjacent to the surface.

\section{Stochastic model} \label{sec_stochastic}

\subsection{Birth-death Markov process}

Rate equation (\ref{eq_rho}) demonstrates the existence
of two steady-state solutions,
but provides no information concerning the rate
at which random fluctuations of the mobile dislocation population
will carry the system past the critical point.
To describe this dynamic behavior,
we model the dynamics as a birth-death Markov process.\cite{gardiner04}
Here the value of $\rho$ can increase or decrease by $\Delta\rho$,
with a transition probability per unit time $\dot{\rho}_+(\rho)/\Delta\rho$
or $\dot{\rho}_-(\rho)/\Delta\rho$,
respectively.
These transitions are independent of the time history of $\rho$,
and correspond to the creation and pinning, respectively,
of one mobile dislocation in one slip plane inside a cell close to the surface.
This behavior can be viewed as a biased random walk along the mobile dislocation
density axis.

For convenience, we define $n = \rho/\Delta\rho$,
so that the step size of every transition is $\pm 1$.
The possible states of the system are thus described by an integer $n$,
which assumes values from 0 to $n_c = \lceil\rho_c/\Delta\rho\rceil$,
where the typical fluctuations are around
$n_* = \lfloor\rho_*/\Delta\rho\rceil$.
Defining $A_2 = a_2 n_c\Delta\rho$ and $B_2 = b_2 n_c\Delta\rho$,
the birth and death rates of the Markov process are
\begin{equation}
  \lambda_n = B_1\sigma^2 e^{\alpha\sigma},\quad
  \mu_n = \frac{B_2 n}{n_c}\sigma \label{eq_rates}
\end{equation}
with $\sigma(n) = A_1 + A_2 n / n_c$.
Using these rates, the rate equation can then be written as
\begin{equation}
  \label{eq_rate_n}
  \dot{n} = \lambda_n(n) - \mu_n(n).
\end{equation}
The corresponding master equation,
describing the time evolution of the probability to be in the state $n$,
is
\begin{equation}
  \frac{\partial P_n(t)}{\partial t} = \lambda_{n - 1} P_{n - 1}(t)
  + \mu_{n + 1} P_{n + 1}(t) - (\lambda_n + \mu_n) P_n(t).
  \label{eq_master_probability}
\end{equation}
Finding the BDR is now equivalent to finding the mean time it takes for the
biased random walker reach $n_c$,
when starting from the vicinity of $n_* = O(1)$.\cite{gardiner04}

\subsection{Estimating the time to breakdown}

\begin{figure}
  \includegraphics[width = 8.5cm]{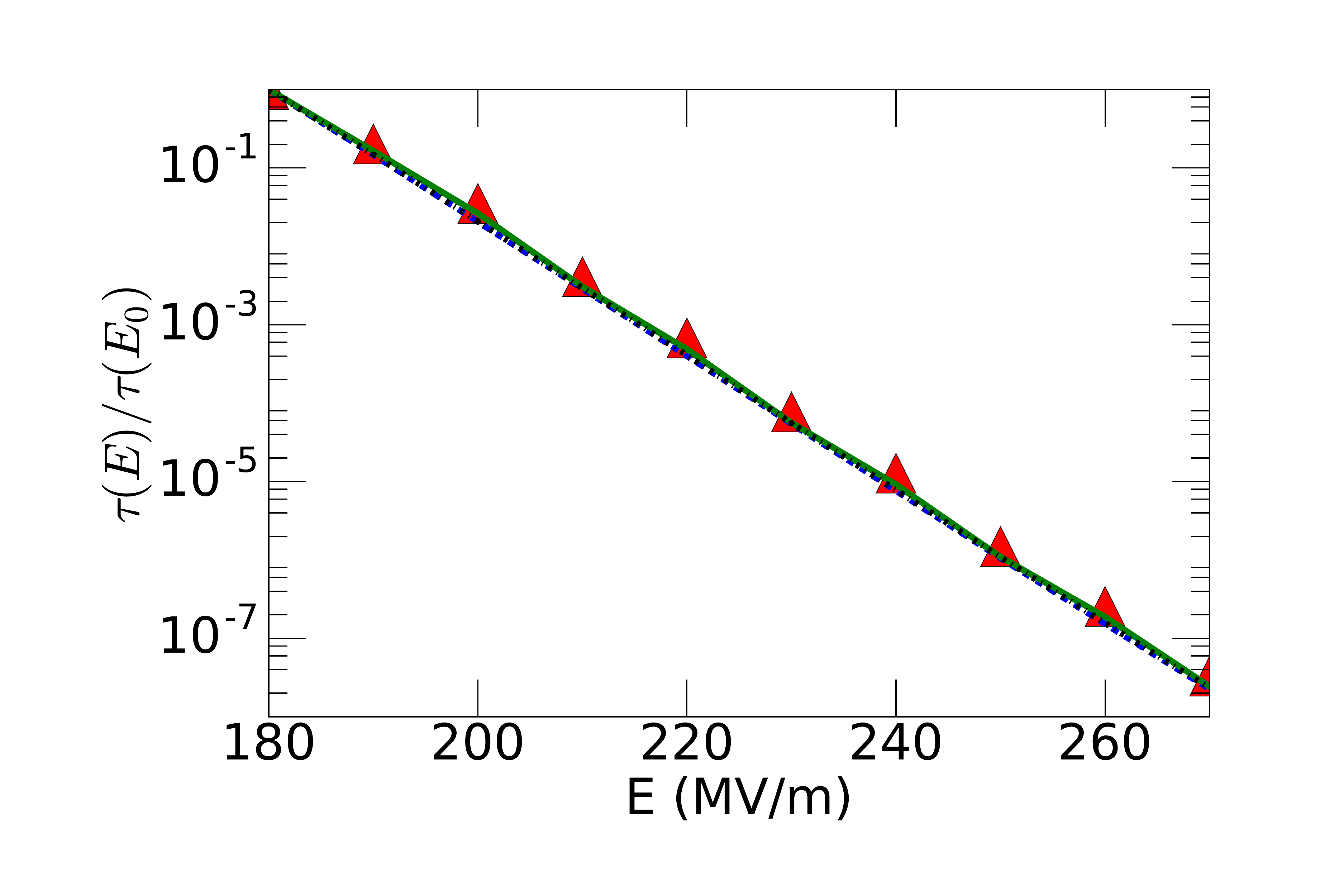}
  \caption
  {
    \label{fig_tau_for_field}
    Mean breakdown time $\tau$ as a function of the electric field
    relative to $\tau(E_0 = 180\,\text{MV/m})$,
    calculated using the exact formula [Eq.~(\ref{eq_tau_exact}), solid line],
    the metastable approximation
    [Eqs.~(\ref{eq_pi_n}) and (\ref{eq_tau_metastable}), dashed line],
    the metastable approximation with an additional Stirling approximation of
    $\pi_n$
    [Eqs.~(\ref{eq_tau_metastable}) and (\ref{eq_tau_metastable_final}),
      dash-dotted line],
    and the simulation (triangles).
    The size of the symbols here and in all symbols accounts for simulation
    errors,
    see Appendix \ref{sec_simulation}.
  }
\end{figure}

Given the values of $\lambda_n$ and $\mu_n$ for every $0 \leq n < n_c$,
the mean time to reach $n_c$ from any state $n$ can be written recursively as
\begin{equation}
  T_n = \frac{\lambda_n}{\lambda_n + \mu_n} T_{n + 1}
  + \frac{\mu_n}{\lambda_n + \mu_n} T_{n - 1} + \frac{1}{\lambda_n + \mu_n}.
  \label{eq_mean_passage_time}
\end{equation}
The solution to this equation,
with the boundary conditions $T_{n_c} = 0$ (absorbing state at $n = n_c$)
and $T_0 = T_1 + \lambda_0^{-1}$ (reflecting boundary at $n = 0$),
is given, for any $n < n_c$, by
\begin{equation}
  T_n = \sum_{i = n}^{n_c}
  \phi_i\left( \sum_{j = 0}^i \frac{1}{\lambda_j \phi_j} \right),
  \label{eq_tau_exact}
\end{equation}
with $\phi_n = \prod_{m = 1}^{n} \mu_m / \lambda_m$.\cite{gardiner04}
Since the system resides in a long-lived metastable state prior to escape,
$T_n$ is independent of $n$, as long as $n = O(n_*)$.
The lines in Fig.~\ref{fig_tau_for_field},
which represent Eq.~(\ref{eq_tau_exact}) and various approximations of it,
see below, agree well with the values found from numerical simulations,
see Appendix \ref{sec_simulation}.

\subsection{Metastable approximation}

The exact solution for $\tau$, presented in the previous section
[see Eq.~(\ref{eq_tau_exact})], is highly cumbersome.
In order to provide insight into the effect of physical
constants and parameters on the BDR,
it is possible to employ a \emph{metastable} approximation (see below).
Starting from some arbitrary initial condition,
the system settles after a relaxation time $t_r$ in a metastable state centered
about $n_*$.
Assuming $t_r \ll \tau$,
we can employ the metastable assumption,
where the probability of being absorbed into $n = n_c$ is given
$P_{n = n_c}(t) = 1 - e^{-t/\tau}$,
while $P_{n < n_c}(t) = \pi_n e^{-t/\tau}$,
where $\pi_n$ is a normalized time-independent
\emph{quasistationary} probability distribution
(QSD).\cite{dykman94, elgart04, assaf06, escudero09, assaf10, assaf17}

\begin{figure}
  \includegraphics[width = 8.5cm]{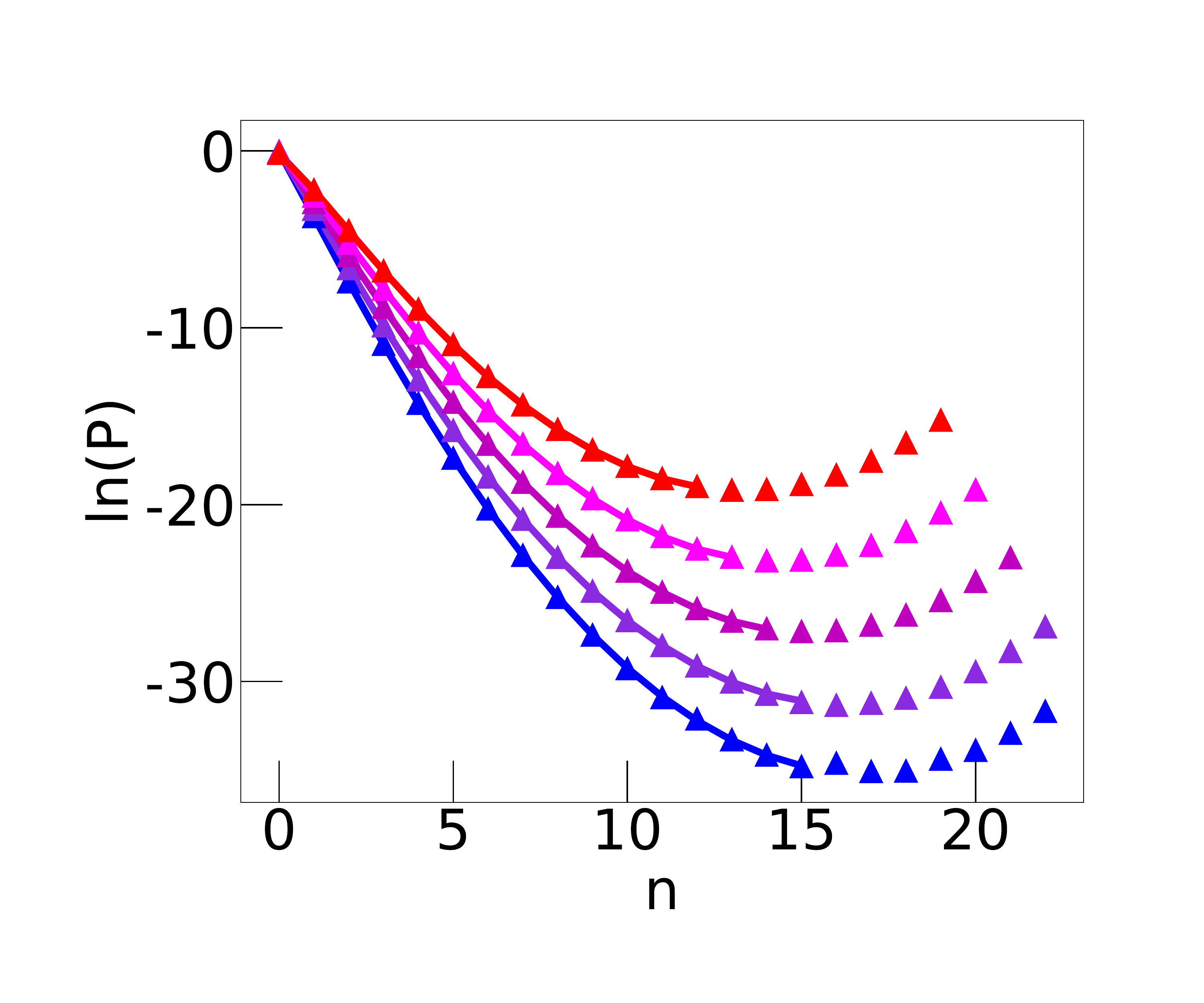}
  \caption
  {
    \label{fig_qsd}
    The probability of being at state $n$,
    calculated from the metastable approximation
    [Eq.~(\ref{eq_pi_n}), line]
    and the simulation (triangles)
    for the nominal parameter set and electric fields of 180,
    200, 220, 240, and 260 MV/m (from bottom to top).
    Here we chose $n_c + 7$ as an absorbing state,
    to clearly demonstrate the minimum at $n = n_c$,
    see Appendix \ref{sec_simulation}.
  }
\end{figure}

Substituting the metastable ansatz into Eq.~(\ref{eq_master_probability})
and assuming that $\tau$ is exponentially large,
to be verified \emph{a posteriori},
yields the \emph{quasistationary} master equation
\begin{equation}
  \lambda_{n - 1}\pi_{n - 1} + \mu_{n + 1}\pi_{n + 1}
  - (\lambda_n + \mu_n)\pi_n = 0.
\end{equation}
Together with the fact that $\mu_0 = 0$ and $\pi_{n < 0} = 0$,
the solution for $\pi_n$ is
\begin{equation}
  \pi_n = \pi_0 \prod_{m = 1}^{n} \frac{\lambda_{m - 1}}{\mu_m},
  \label{eq_qsd}
\end{equation}
where $\pi_0$ is found via the normalization condition
$\sum_{n = 0}^{n_c - 1}\pi_n = 1$.\cite{gardiner04}
Substituting the values of $\lambda_n$ and $\mu_n$
from Eq.~(\ref{eq_rates}) into Eq.~(\ref{eq_qsd}) yields
\begin{align}
  \pi_n &= \pi_0\frac{n_c\eta}{n + n_c\eta}
  \left(\frac{A_1 B_1}{B_2\eta}\right)^n \label{eq_pi_n} \\
  &\times \exp\left[n\alpha A_1\left(1 + \frac{n - 1}{2n_c\eta}\right)\right]
  \frac{\Gamma(n + n_c\eta)}{n!\,\Gamma(n_c\eta)}, \nonumber
\end{align}
where $\eta = A_1/A_2$, and $\Gamma(x)$ is the Gamma function.

For the nominal set of parameters, up to the close vicinity of $E = E_b$,
$\pi_0 \gg \pi_{n > 0}$, and therefore $\pi_0 \approx 1$.
Alternatively, to achieve a more accurate normalization of the distribution,
we notice that the maximum of the distribution is obtained at $n = 0$,
and the width of the distribution is $O(1)$.
As a result,
the bulk of the QSD can be found by linearizing the reaction rates
close to the maximum,
and obtaining $\lambda_n \approx A_1^2 B_1 e^{\alpha A_1}$ and
$\mu_n \approx A_1 B_2 n / n_c$.
Using Eq.~(\ref{eq_qsd}),
the approximate QSD resulting from these linear rates
is a Poisson distribution with a mean (and variance) of
\begin{equation}
  \mathcal{R} = \frac{A_1 B_1}{B_2} n_c e^{\alpha A_1}. \label{eq_poisson_mean}
\end{equation}
Therefore, the normalization factor for the QSD is $\pi_0 = e^{-\mathcal{R}}$.
Figure \ref{fig_qsd} shows excellent agreement between the theoretical
and simulation results for $\pi_n$,
for the nominal parameters and five different electric fields.

Since the flux through $n_c$ determines the escape rate,
the mean breakdown time is given by
\begin{equation}
  \tau \simeq (\lambda_{n_c}\pi_{n_c})^{-1}, \label{eq_tau_metastable}
\end{equation}
which is found from Eq.~(\ref{eq_master_probability}) for $n = n_c$.

Using the Stirling approximation $\Gamma(z) = (2\pi/z)^{1/2}(z/e)^z$,
the rightmost factor in Eq.~(\ref{eq_pi_n}), containing the Gamma functions,
becomes, for $n = n_c$,
\begin{equation}
  \frac{\Gamma(n_c + n_c\eta)}{n_c!\,\Gamma(n_c\eta)} =
  \frac{1}{\sqrt{2\pi n_c}}(1 + \eta)^{n_c}
  \left(1 + \frac{1}{\eta}\right)^{n_c\eta - 1/2}.
\end{equation}
Therefore,
\begin{equation}
  \pi_{n_c} =
  \frac{e^{-\mathcal{R}}}{\sqrt{2\pi n_c}}
  \left(1 + \frac{1}{\eta}\right)^{-3/2}
  \exp\left(-\frac{\alpha A_1}{2\eta}\right)
  e^{-n_c\Delta S},
\end{equation}
where
\begin{equation}
  \label{eq_term_in_exponent}
  \Delta S = \ln\frac{B_2}{A_1 B_1} -
  \alpha A_1 \left(1 + \frac{1}{2\eta}\right) - (\eta + 1)
  \ln\left(1 + \frac{1}{\eta}\right).
\end{equation}
Plugging this result, together with $\lambda_{n_c}$,
into Eq.~(\ref{eq_tau_metastable}), yields
\begin{equation}
  \tau = \mathcal{A} e^{n_c \Delta S}, \label{eq_tau_metastable_final}
\end{equation}
with
\begin{equation}
  \label{eq_prefactor}
  \mathcal{A} = \sqrt{2\pi n_c}
  \frac{\exp\left[\mathcal{R} - \alpha A_1
      \left(1 + \frac{1}{2\eta}\right)\right]}{A_1^2 B_1}
  \left(1 + \frac{1}{\eta}\right)^{-1/2}.
\end{equation}
Here,
$n_c\Delta S$ serves as a barrier that the system needs to overcome
in order to enter the breakdown state.

As an alternative to the discrete calculation in Eq.~(\ref{eq_qsd}),
it is possible to employ the WKB ansatz, and express $\pi_n$ as an exponential
function\cite{dykman94, escudero09, assaf10, assaf17}
\begin{equation}
  \pi(q) \sim \exp\{n_c[S(q_*) - S(q)] + S_1(q_*) - S_1(q)\}, \label{eq_pi_wkb}
\end{equation}
where $q = n/n_c$, $q_* = n_*/n_c$, and
\begin{equation}
  S(q) = -\int^{q}\ln\frac{w_+(\xi)}{w_-(\xi)}d\xi,\quad
  S_1(q) = \frac{1}{2}\ln[w_+(q)w_-(q)].
\end{equation}
Here $w_+(q) = \lambda(n_c q)/n_c$ and $w_-(q) = \lambda(n_c q)/n_c$.
Although the WKB approximation is not formally valid when
$n_* = O(1)$ as in our case,\cite{escudero09, assaf10, assaf17}
since the barrier for breakdown is large,
using Eq.~(\ref{eq_pi_wkb}) to calculate the QSD and $\tau$,
for various electric fields,
yields results which coincide in the leading order
with those of the above method.\cite{beer2015}

\begin{figure}
  \centering
  \includegraphics[width=8.5cm]{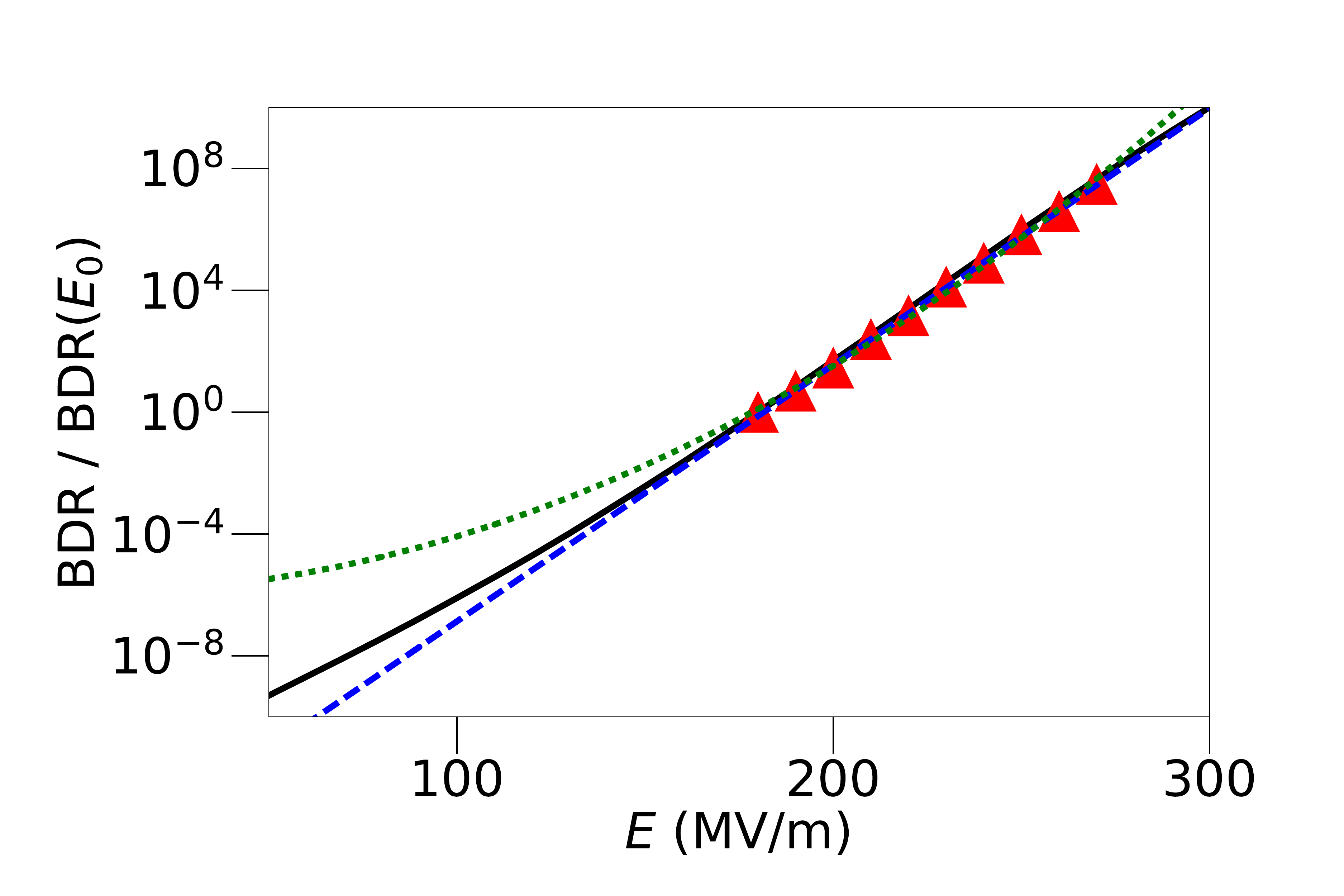}
  \caption
  {
    \label{fig_linear_vs_square}
    BDR as a function of the electric field.
    The solid line is the metastable approximation
    [Eq.~(9) in the main text],
    the triangles are the simulation results,
    and the dashed and dotted lines are linear and quadratic fits,
    respectively, see text.
  }
\end{figure}

Our analytical results,
given by Eqs.~(\ref{eq_term_in_exponent})-(\ref{eq_prefactor}),
contain a nontrivial dependence of $\tau$ on $E$.
Indeed,
while it can be shown that for $E \simeq E_c$ the term
$A_1 \sim E^2$ dominates the exponent in Eq.~(\ref{eq_tau_metastable_final}),
for $E < E_c$, where breakdown is fluctuation-driven,
our results can be approximated by a linear dependence of $\ln\tau$ on $E$,
\begin{equation}
  \tau \simeq \mathcal{C}\exp[\gamma\left(1 - E / E_0)\right]. \label{eq_linear_ln_tau}
\end{equation}
Here $E_0$ is a reference field,
and $\gamma$ and $\mathcal{C}$ are constants independent of $E$.
This is demonstrated in Figs.~\ref{fig_tau_for_field} and
\ref{fig_linear_vs_square} (for fields between 50 and 300 MV/m).
Note that, while within the range of currently available data,
this behavior is similar to that derived in Ref.~\onlinecite{nordlund12},
$\tau \sim \exp(\alpha E^2)$,
the models diverge outside that range, see Fig.~\ref{fig_linear_vs_square}.

\section{Model fitting and validation} \label{sec_model}

As described in Section \ref{sec_mean},
there are four parameters in the MDDF model whose values are not taken
from standard properties of the cathode material
or estimated from direct observations.
The first two of these are the free energy of activation $E_a$
and the activation volume $\Omega$
of mobile dislocation nucleation in Eq.~(\ref{eq_drhop}).
The third parameter is $\kappa$, see Eq.~(\ref{eq_drhop}),
which is a temperature-independent kinetic prefactor of the rate
constant of dislocation nucleation.
Finally, the fourth parameter $\beta$, in Eq.~(\ref{eq_stress}),
represents the in-plane effective attenuation or enhancement
of the electromagnetic field.

The purpose of this section is to describe the calibration
of these four parameters by fitting the results of the model
to experimental data of BDRs
as a function of the electric field and the temperature.
The quality of the fit can serve as a validation of the model,
and the resulting values will be compared to previous estimates,
and used to predict the results of future experiments.

Most of the available experimental data was acquired from the
CLIC prospect study,
in which the BDRs are measured in breakdowns per pulse
per meter of accelerator (bpp/m).
Thus,
to translate bpp/m units to the natural characteristic time of the MDDF model,
the mean breakdown time per slip plane $\tau$,
the CLIC accelerator geometry must be taken into consideration.
Every meter of the CERN CLIC accelerator is planned to contain
100 cathode irises,
in each of which a ring of 2.35 mm diameter and 1 mm width
is subjected to the electric field pulse.\cite{clic_stage}
The surface area of one dislocation cell is approximately
$\Delta\rho^{-2} = 10^{-4}$ mm.
Assuming an active slip plane can develop independently in each
dislocation cell,
the number of active slip planes in one meter of accelerator length is then
$N \approx 1.5\times 10^7$.
The BDR, in bpp/m, is $tN/\tau$, where $t$ is the time duration of one pulse.
With $t$ = 230 ns in the experimental data,\cite{grudiev09}
the resulting conversion of units between the BDR and $\tau$ is
\begin{equation}
  \label{eq_conversion_ratio}
  R = 3.45\,\frac{\text{bpp}\,\text{s}}{\text{m}}\,\tau^{-1}.
\end{equation}

\begin{figure}
  \includegraphics[width = 8.5cm]{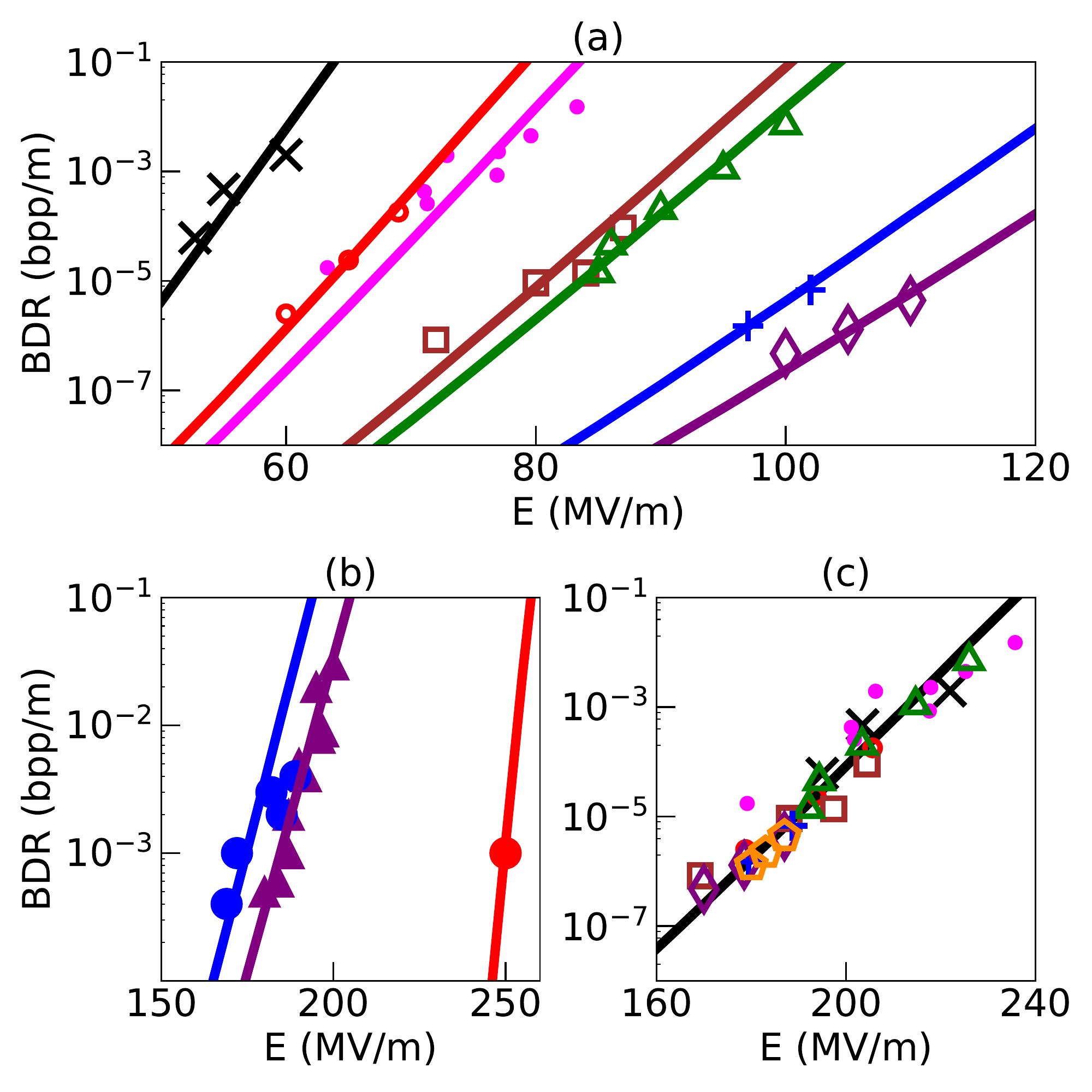}
  \caption
  {
    \label{fig_fits}
    BDR as a function the electric field:
    (a) Group (1) in the text, measured at 300 K in various
    structures.\cite{grudiev09}
    (b) Group (2) in the text.
    The two data sets on the left were measured at 300 K,
    and the set on the right was measured at 45 K.\cite{cahill17}
    In Figs. (a) and (b),
    the lines represent a fit to the MDDF model
    with the nominal set of parameters,
    except for $\beta$, which varies among the data sets.
    (c) Group (3) in the text represented by the yellow
    pentagons,\cite{wuensch17}
    and group (1), with the electric fields scaled,
    represented by all other symbols.
    The line in Fig.~(c) represents a fit to the MDDF model with the nominal
    set of parameters (including $\beta$).
  }
\end{figure}

The numerical results of the MDDF model can be compared to experimental data,
which consists of sets of measurements
of the BDR as a function of the electric field,
where each set of measurements was taken in a different physical structure.
The sets are divided into three groups:
(1) Seven sets measured in different structures in the CERN CLIC project
at room temperature [Fig.~\ref{fig_fits}(a)].\cite{grudiev09}
(2) Three sets measured at SLAC [Fig.~\ref{fig_fits}(b)].
The two sets on the left of the figure were measured at room temperature
(300 K),
and the set on the right was measured at 45 K.
The leftmost set and the set on the right were measured
in the same structure.\cite{cahill17}
Although the data set at 45 K consists of a number of measurements,
the field varies over a small range.
Thus, we consider this data as a single average value.
(3) A set measured in the CERN CLIC project
[the pentagons in Fig.~\ref{fig_fits}(c)].
This set of data is considered by CLIC to be the most accurate
to date,\cite{wuensch17}
and will therefore be used here as a reference set.

\begin{figure}
  \centering
  \includegraphics[width = 8.5cm]{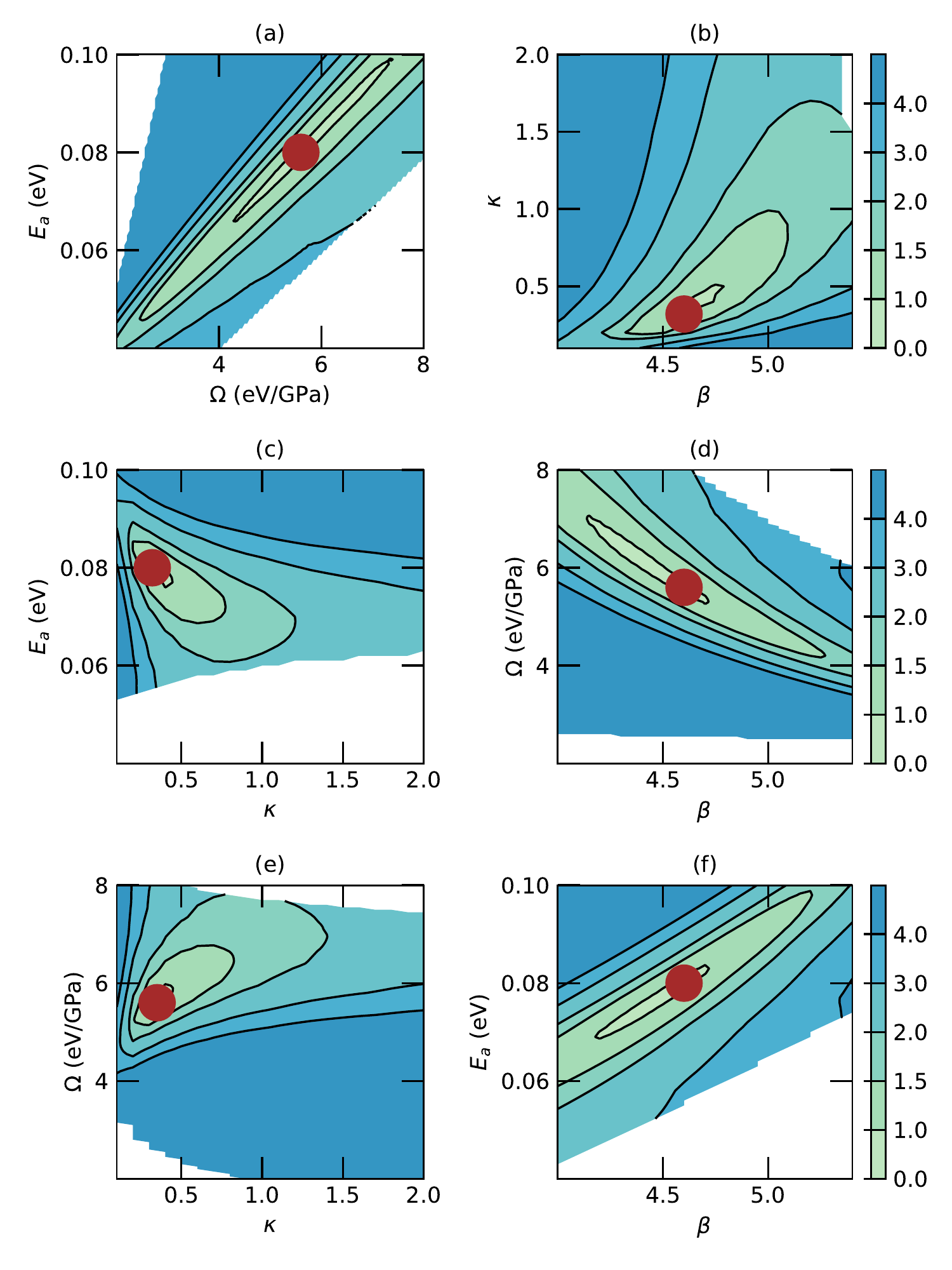}
  \caption
  {
    \label{fig_colormaps}
    The LSQ fit measure $Q$ [see Eq.~(\ref{eq_lsq})] as a function of
    (a) $\Omega$ and $E_a$,
    (b) $\beta$ and $\kappa$, (c) $\kappa$ and $E_a$, (d) $\beta$ and $\Omega$,
    (e) $\kappa$ and $\Omega$, and (f) $\beta$ and $E_a$.
    In each graph,
    the two remaining parameters of the set ($E_a$, $\Omega$, $\kappa$, $\beta$)
    are held at their nominal values.
    The circle shows the location of the nominal parameter set
    in the phase space.
  }
\end{figure}

Since the parameters $E_a$, $\Omega$,
and $\kappa$ should depend on the properties of the material itself,
which is identical for all structures,
we expect the value of $\beta$ alone to vary among the structures.
Despite the fact that $\beta$ is not known \emph{a priori} for any structure,
it is clear from Eq.~(\ref{eq_stress})
that the MDDF model is invariant for a constant $\beta E$.
Therefore, rescaling the electric field should yield a fit,
for all the data sets, with the same values for all four parameters.
Figure \ref{fig_fits}(c) shows the data sets from group (1) rescaled so that
their measured BDRs as a function of the field are all fitted
by the same exponential relation as that of the reference set.
A LSQ fit to the rescaled data was performed
(see Appendix \ref{sec_fit} for details),
yielding an optimal fit for the nominal set described
in Section \ref{sec_mean},
namely $E_a$ = 0.08$\pm$0.01 eV, $\Omega$ = 5.6$\pm$0.1 eV/GPa,
$\kappa$ = 0.32$\pm$0.01, and $\beta$ = 4.6$\pm$0.1.
Figure \ref{fig_colormaps} shows the value of the quality measure of the fit,
$Q$,
for the six two-dimensional cross sections of the four-dimensional phase space
$(E_a, \Omega, \kappa, \beta)$.
The circle shows the location of the nominal parameter set in each
cross section.

The activation energy $E_a$ = 0.08 eV is consistent with that
previously found for dislocation nucleation from existing sources,\cite{zhu08}
and considerably lower than the activation energy for
dislocation nucleation in configurations with no preexisting
sources.\cite{bonneville88, couteau11, ryu11}
The activation volume is $\Omega$ = 5.6 eV/GPa = 57$b^3$,
with $b$ the Burgers vector.
This result is consistent with experimental results,
in which the activation volume was found to be within the range
$10b^3 < \Omega < 124b^3$.\cite{zhu08, couteau11}

\section{Sensitivity of the model to physical assumptions}
\label{sec_sensitivity}

In this section we consider possible variations of the physical model,
and examine the effect they would have on the predictions of the MDDF model.

\subsection{Dependence of stress on dislocation density}

The MDDF model discusses in-plane mobile dislocation density fluctuations,
neglecting interactions between slip planes.
The mobile dislocation density $\rho$ is therefore
a two-dimensional density,
measured in units of length per area, nm\textsuperscript{-1}.
In the case where $\rho$ is defined as the volume density of mobile dislocations
in units of length per volume,
nm\textsuperscript{-2},
the average distance between dislocations is propotional
to $\rho^{-1/2}$.\cite{taylor34, progress80}
The stress in this case is
$\sigma = \epsilon_0 (\beta E)^2 / 2 + ZGb\rho^{1/2}$,
leading to modified creation and depletion rates
[Eqs.~(\ref{eq_drhop}) and (\ref{eq_drhom})] of
\begin{equation}
  \dot{\rho}^+ = \frac{25\kappa C_t c}{G^2 b} \sigma^2
  \exp\left(-\frac{E_a - \Omega\sigma}{k_B T}\right),\quad
  \dot{\rho}^- = \frac{50\xi C_t c}{G} \sigma b\rho
\end{equation}
where $c$ = 1 $\mu$m\textsuperscript{-2} is now
the volume density of the barriers,
while all other constants retain their original meaning.
The factor of $b$ in the depletion term was added in order to correctly describe
the probability of two dislocations interacting,
now in a volume instead of a plane,
assuming that the width of a dislocation is equal to the Burgers vector $b$.

\begin{figure}
  \centering
  \includegraphics[width=8.5cm]{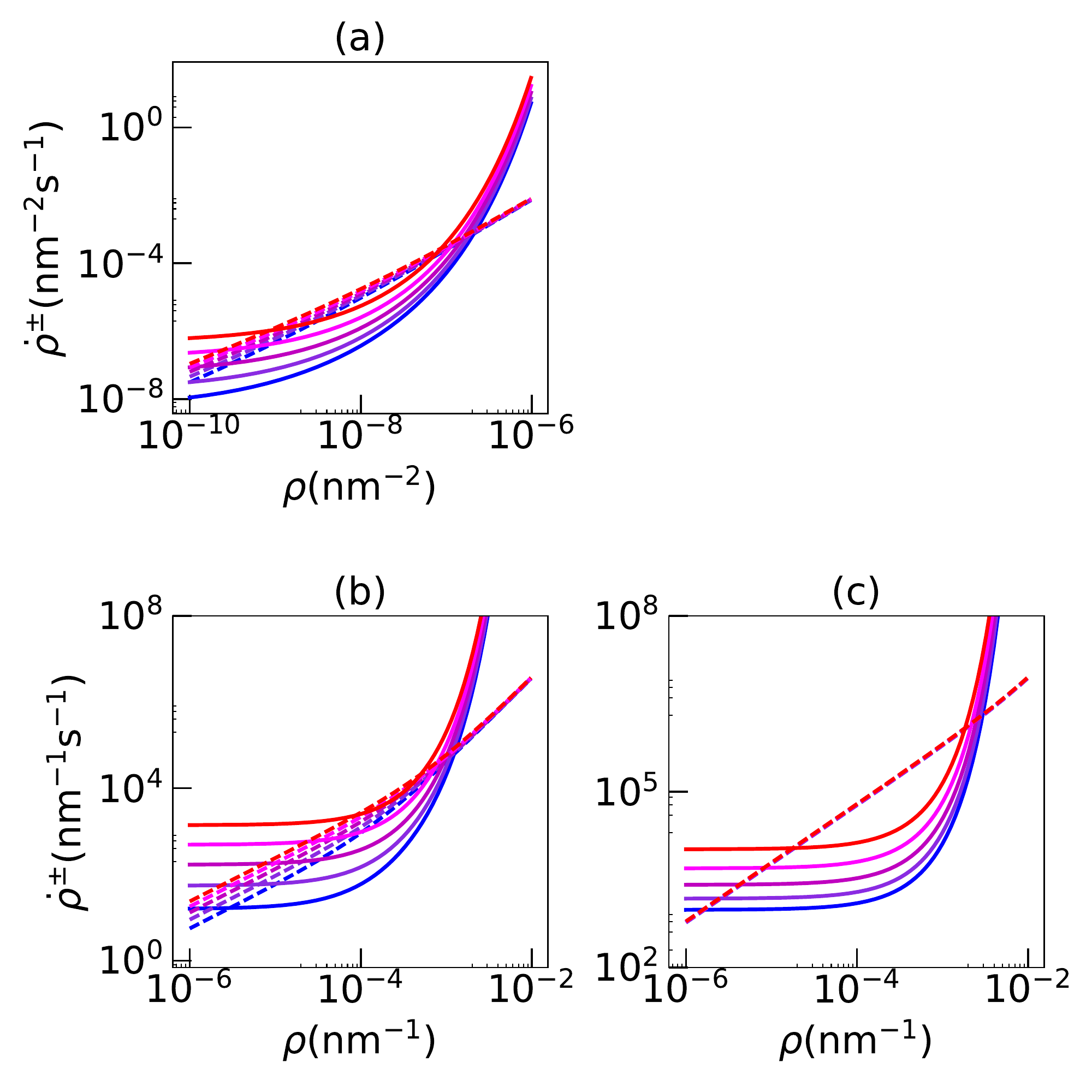}
  \caption
  {
    \label{fig_alternative_models}
    $\dot{\rho}^+$ (solid lines) and $\dot{\rho}^-$ (dashed lines)
    for the nominal set of parameters and an electric field of
    (from bottom to top) 150, 190, 230, 270, and 310 MV/m,
    in
    (a) a model describing mobile dislocation volume density fluctuations,
    (b) a model with $\sigma \sim \rho^2$ (here $Z = 5\times 10^4$),
    and (c) a model in which both sessile and mobile dislocations
    contribute to the average stress.
  }
\end{figure}

As seen in Fig.~\ref{fig_alternative_models}(a),
for adjusted values of the parameter set
$E_a$, $\Omega$, $\kappa$, and $\beta$,
the volume density creation and depletion rates,
$\dot{\rho}^+$ and $\dot{\rho}^-$,
exhibit the same qualitative behavior as in the two-dimensional density model.
The same considerations as in the latter model can then be applied,
once again yielding the $\ln\tau \sim E$ dependence described
in Section \ref{sec_stochastic}.

In general,
the stress can have a power dependence on the mobile dislocation density
of the form $\sigma = \epsilon_0 (\beta E)^2 / 2 + ZG(b\rho)^\nu$
with some value of $\nu$.
When considering a volume dislocation density we took $\nu = 1/2$,
with an additional correction to Eq.~(\ref{eq_drhom})
due to dimensional considerations.
As another example, we examine the case in which $\nu = 2$,
i.e.,
the stress is proportional to the two-dimensional dislocation density squared.
Here, the value of $Z$,
the proportionality constant linking the stress to the mobile dislocation
density,
is expected to be different.
Indeed, choosing $Z = 5\times 10^4$,
the same qualitative behavior of $\dot{\rho}^+$ and $\dot{\rho}^-$ can
be produced for the nominal parameters found in Section \ref{sec_model},
as can be seen in Fig.~\ref{fig_alternative_models}(b).

\subsection{Effect of sessile and mobile dislocations}

Another assumption in the model, justified in Section \ref{sec_mean},
is that the stress is affected by the mobile dislocations only.
If the stress caused by sessile dislocations
contributes significantly to the overall stress,
Eq.~(\ref{eq_stress}) becomes
$\sigma = \epsilon_0 (\beta E)^2 / 2 + Z_m Gb\rho + Z_s Gbs$,
with $Z_m$ and $Z_s$ being proportional factors defining the relative
contributions of the mobile and sessile dislocations to the stress,
respectively,
and $s \approx 20\,\mu\text{m}^{-1}$ the density of sessile dislocations
(see Section \ref{sec_mean} and Fig.~\ref{fig_sessile_dislocations}).
Examining the extreme case in which $Z_m = Z_s = 1$,
a parameter set can be found for which $\dot{\rho}^+$ and $\dot{\rho}^-$
exhibit the same qualitative behavior as in the original model,
where $Z_s = 0$, as seen in Fig.~\ref{fig_alternative_models}(c).
Here, too,
the calculation yields a $\ln\tau \sim E$ dependence
as in Section \ref{sec_stochastic}.

\section{Proposed experiments} \label{sec_proposed}

\subsection{Temperature dependence}

\begin{figure}
  \centering
  \includegraphics[width = 8.5cm]{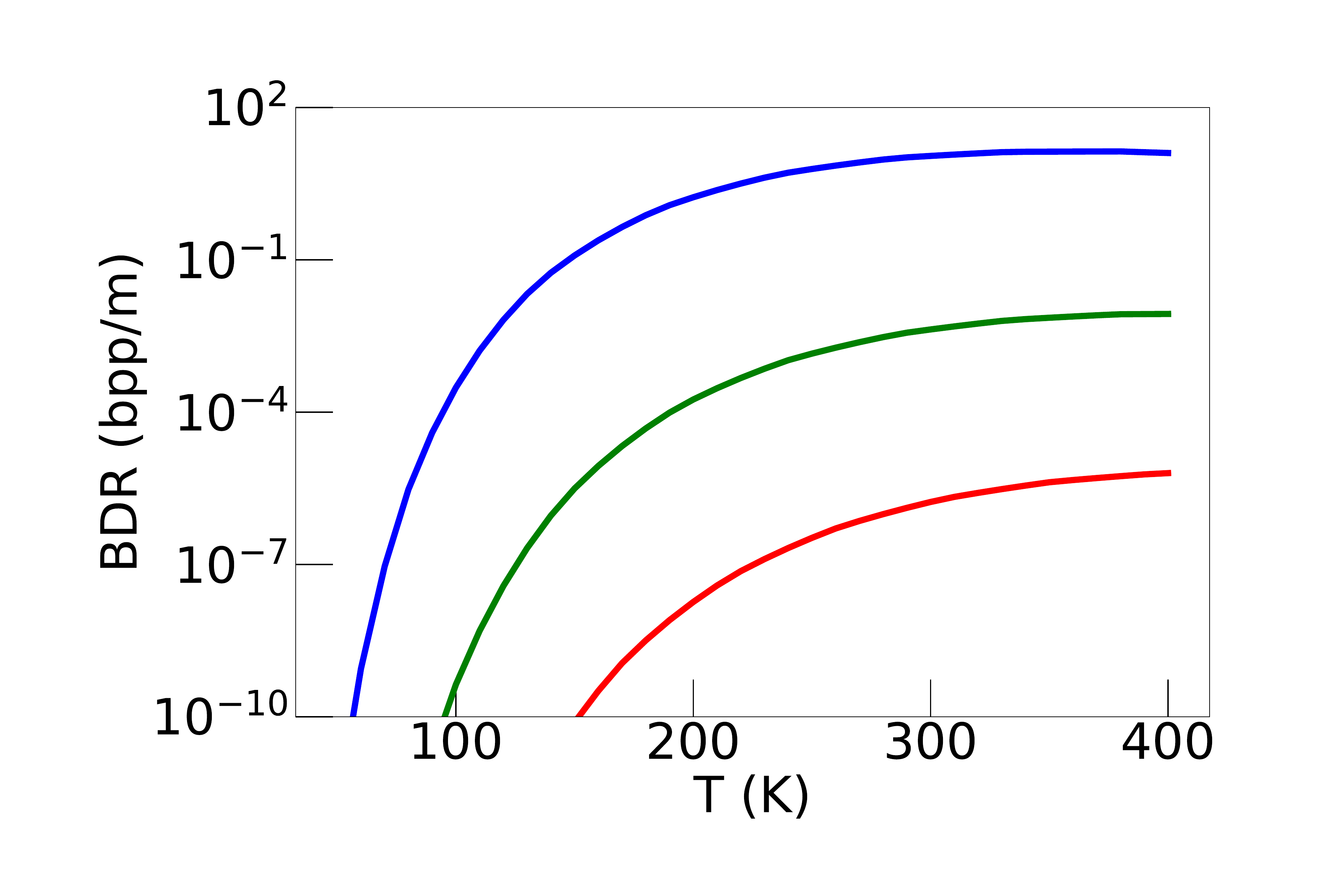}
  \caption
  {
    \label{fig_tau_for_temperature}
    BDR as a function of the temperature for the nominal parameter set,
    calculated using the metastable approximation
    [Eq.~(\ref{eq_tau_metastable_final})].
    The lines, from bottom to top, are for fields of 180, 220, and 260 MV/m.
  }
\end{figure}

\begin{figure}
  \centering
  \includegraphics[width = 8.5cm]{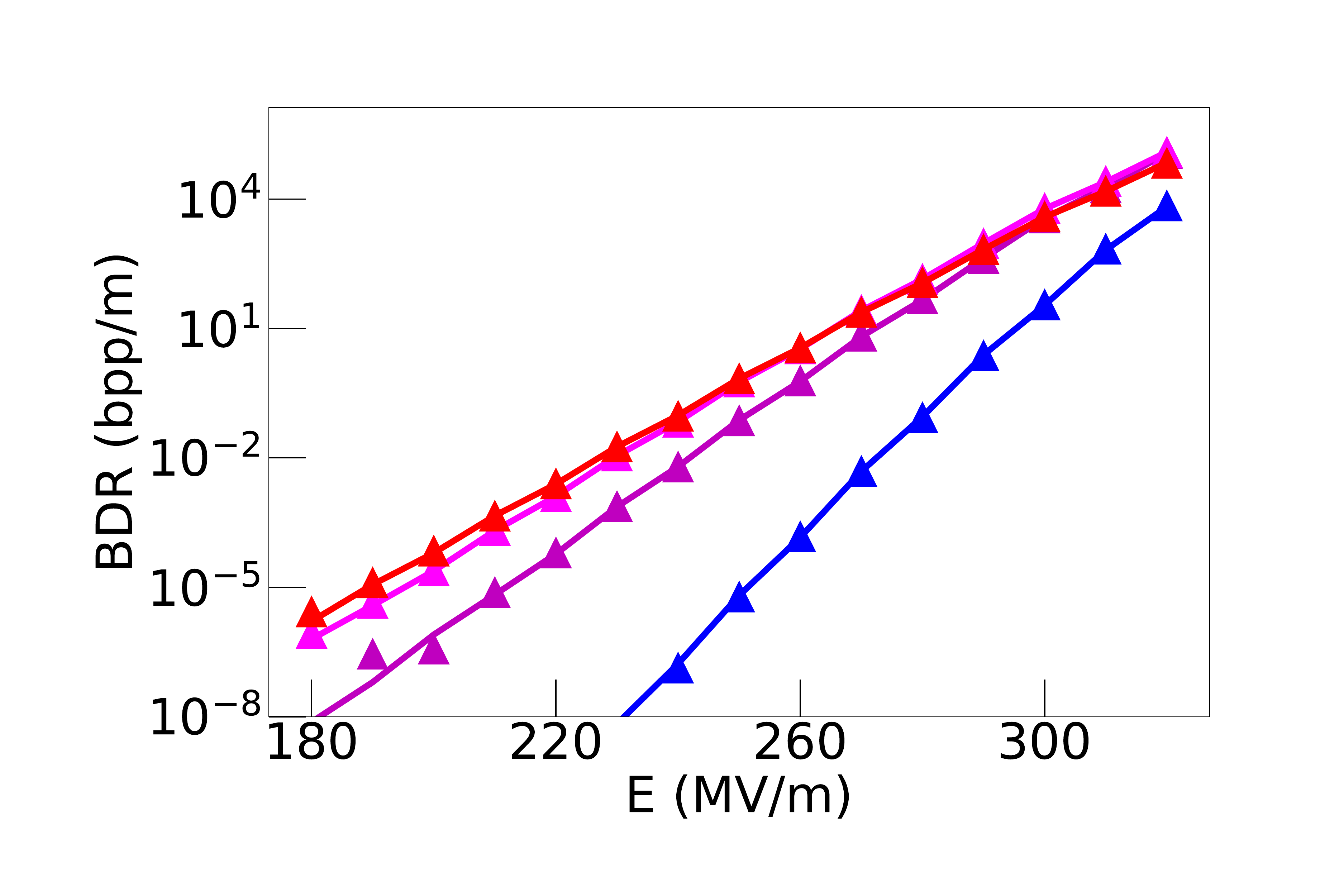}
  \caption
  {
    \label{fig_fields_in_temperatures}
    BDR as a function of the electric field calculated using the exact formula
    [Eq.~(\ref{eq_tau_exact}), lines]
    and the simulation (triangles).
    The results are plotted, from bottom to top,
    for temperatures of 100, 200, 300, and 400 K.
  }
\end{figure}

As discussed in Section \ref{sec_model},
experiments to date were carried out primarily at a temperature of 300 K.
In calibrating the model to find the unknown parameters,
only one measurement point at a different temperature of 45 K
was available.\cite{cahill17}
However,
the model predicts a strong dependence of the BDR on temperature,
due to the explicit dependence of $\alpha$ in
Eq.~(\ref{eq_tau_metastable_final}) on the temperature,
and the fact that $n_c$ decreases for increasing temperature.
Figure \ref{fig_tau_for_temperature} shows the dependence of the BDR
on the temperature for three electric fields,
and Fig.~\ref{fig_fields_in_temperatures} shows the BDR as a function of the
electric field for four different temperatures.
The effect of the temperature is the most pronounced for weaker electric fields,
because the stronger the electric field is,
the greater the stress and therefore the lower the activation enthalpy is,
thus making the temperature less significant in Eq.~(\ref{eq_drhop}).

Experiments, supplying data of BDRs at different temperatures and fields,
would be instrumental for determining the nature of the temperature dependence
of the BDR.
This dependence can then be compared to the predictions of the model,
and can be used, in addition,
to produce more accurate estimates of the activation energy and volume
of dislocation nucleation.

\subsection{Pulse length dependence}

As mentioned in Section \ref{sec_introduction},
the electromagnetic field driving the breakdown, in some applications,
is a pulsed RF signal.
In the context of the CLIC project, for example,
pulse lengths $t_\text{pulse}$ between 50 and 400 ns were examined,
with a duty cycle of 50 Hz.\cite{wuensch13}

\begin{figure}
  \centering
  \includegraphics[width = 8.5cm]{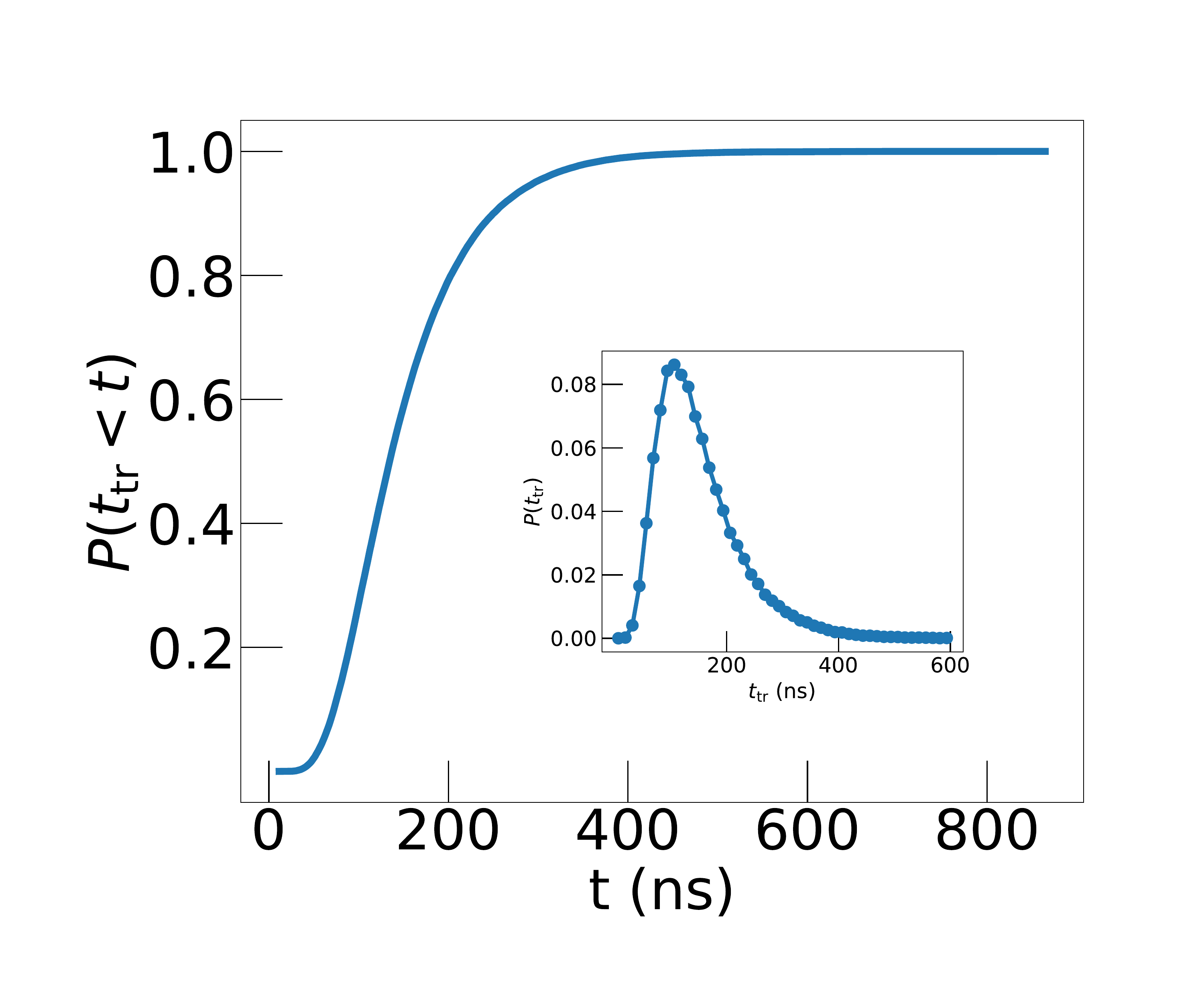}
  \caption
  {
    \label{fig_cdf}
    Cumulative distribution function of the time duration of the trajectory to
    the critical point $t_\text{tr}$ (see text),
    for the nominal set of parameters and an electric field of 250 MV/m.
    The inset shows the probability distribution function,
    drawn as a histogram of forty-eight bins, each bin 12.5 ns wide.
    Both curves were found by simulating $10^5$ breakdown events.
  }
\end{figure}

\begin{figure}
  \centering
  \includegraphics[width = 8.5cm]{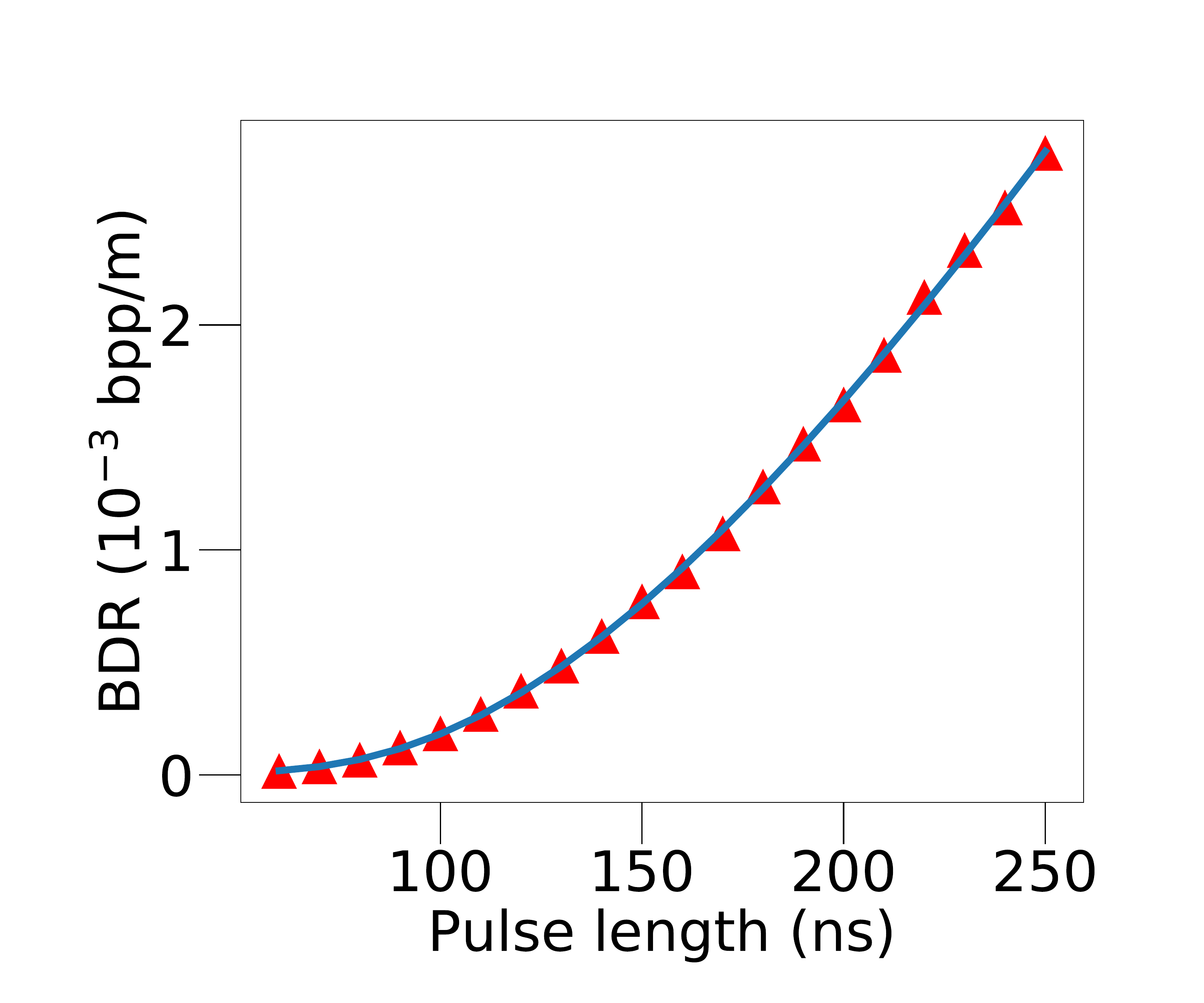}
  \caption
  {
    \label{fig_bdr}
    BDR as a function of the pulse length, $t_\text{pulse}$,
    for the nominal set of parameters and an electric field of 250 MV/m,
    found from the numerical simulations (triangles).
    The line is a fit to Eq.~(\ref{eq_bdr_for_pulse_length}).
  }
\end{figure}

Figure \ref{fig_cdf} shows the cumulative probability distribution function
(CDF) of $t_\text{tr}$ in a simulation,
with the nominal set of parameters and $E$ = 250 MV/m.
Here $t_\text{tr}$ is defined as the time it takes to reach the critical point
$n = n_c(E)$,
starting from $n = n_*(E)$.
For $t_\text{pulse}$ on the order of $t_\text{tr}$ (or shorter),
a significant number of trajectories,
which would have reached the critical point,
will rapidly go to $n = 0$ once the field is switched off.
Therefore, in this regime,
we expect a strong dependence of the BDR on $t_\text{pulse}$,
which can be empirically shown to satisfy
\begin{equation}
  R = R_0 + m(t_\text{pulse} - t_0)e^{-\delta / t}, \label{eq_bdr_for_pulse_length}
\end{equation}
see Fig.~\ref{fig_bdr}.
Here, $R$ is the BDR, and $R_0$, $t_0$, $m$,
and $\delta$ are constants depending on the field.

The BDR was shown, experimentally,
to have an exponential or power-law dependence on
$t_\text{pulse}$.\cite{dobert95, grudiev09, degiovanni16}
However,
the validity of using the existing data to determine the dependence is limited,
as it consists of either a small sample,\cite{dobert95}
or of measurements taken during the conditioning process,
when the BDR is still dominated by extrinsic processes.\cite{degiovanni16}
At the very least,
the BDR is expected to saturate for a continuous-wave RF signal,
and therefore the exponential or power-law dependence holds only for
a limited range of pulse lengths.

\begin{figure}
  \centering
  \includegraphics[width = 8.5cm]{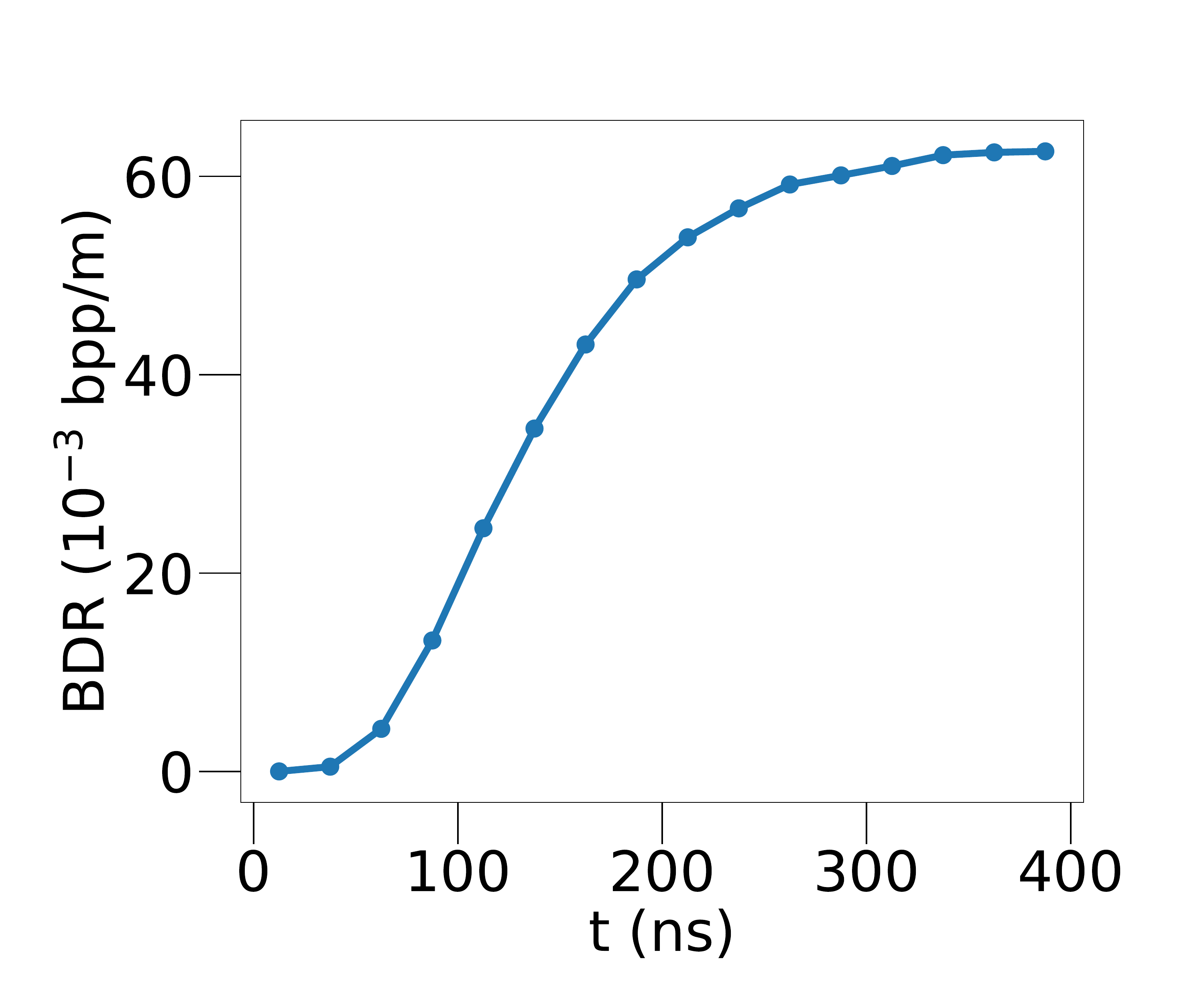}
  \caption
  {
    \label{fig_bd_distribution_within_pulse}
    Probability of a breakdown occuring as a function of the time
    within the pulse,
    found by simulation,
    for the nominal set of parameters, an electric field of 250 MV/m,
    and a total pulse duration of 400 ns.
    The probability distribution is presented as a histogram of sixteen bins,
    each bin 25 ns wide,
    and normalized by the total BDR at 250 MV/m and a pulse length of 400 ns.
    The total number of breakdowns is $10^6$.
  }
\end{figure}

Under the assumption of pulse independence,
the distribution of breakdowns in time within each pulse should be
an increasing function,
due to the finite evolution time $t_\text{tr}$.
Figure \ref{fig_bd_distribution_within_pulse} shows this
distribution for $E$ = 250 MV/m and a pulse duration of 400 ns.
For a given time interval $(t, t + dt)$ within a pulse,
an event will mature if it began
within the time interval $(t - t_\text{tr}, t + dt - t_\text{tr})$.
Given that a breakdown occurred,
the probability that it occurred within an interval $dt$ is,
therefore, $(dt/t_\text{pulse})\int_0^t P(t_\text{tr})dt_\text{tr}$,
where $P(t_\text{tr})$ is linearly proportional to the
probability distribution function shown in the inset of Fig.~\ref{fig_cdf}.
This integral, however, is simply the CDF of $t_\text{tr}$,
and therefore the probability distribution in
Fig.~\ref{fig_bd_distribution_within_pulse} is linearly proportional to the
CDF in Fig.~\ref{fig_cdf}.
This non-Poissonian distribution becomes predominantly Poissonian for times
that are significantly greater than $t_\text{tr}$.

If, however,
the interval between pulses is smaller than the typical relaxation time,
then the breakdown probability should not depend on the pulse duration alone,
but rather on the combined effect of exposure to the field
and the relaxation achieved between pulses.
In this case,
the variation in the BDR within the pulse can be small,
and characterized by a constant probability,
similarly to the slow variation observed for $t > 500$ ns in Fig.~\ref{fig_cdf}.
Indeed, in Ref.~\onlinecite{wu17} it was shown that the breakdown distribution
does not vary significantly within the pulse.
However,
an increase in the breakdown probability was observed for one of the
structures studied in Ref.~\onlinecite{wu17}.
Due to this fact, together with the need to correct for conditioning effects,
we chose not to include this data as a constraint on the MDDF model.
As explained, if the system does not reach full relaxation between pulses,
we expect the BDR to depend on the duty cycle of the pulses,
rather than solely on the pulse length.
Therefore, experiments involving variation of the duty cycle,
as well as further data regarding the pulse-length dependence
during and after conditioning,
can help determine the exact nature of the pulse-length
dependence of the BDR.
This may serve to quantify the memory effect between and within pulses.

\subsection{Field ramping}

Traditionally,
kinetic processes leading to transitions have been studied by varying the
driving force at a constant rate,
measuring changes in the observed transition rate.\cite{blaine12}
In general,
increasing the electric field at a constant rate $\chi$ leads to a
corresponding mean breakdown field $E_{\text{BD}}(\chi)$.
Using the $\tau(E)$ dependence from Eq.~(\ref{eq_linear_ln_tau}) for
constant fields,
we find that, if the field at time $t$ is $E$,
the upper limit of the mean breakdown time is $t + \tau\bm{(}E(t)\bm{)}$.
Then, from Eq.~(\ref{eq_linear_ln_tau}) and the relation $t = E / \chi$,
the upper limit of the mean field at which breakdown occurs is
$E + \chi\mathcal{C}\exp[\gamma\left(1 - E / E_0)\right]$.
Assuming an adiabatic increase of $E$, i.e., $\chi\tau \ll E$,
we can use this upper limit as an estimate of the mean breakdown field.
The lowest upper limit fulfilling this condition for any $E$ is
\begin{equation}
  E_\text{BD} = \frac{E_0}{\gamma}\left( \gamma -
  \ln\frac{E_0}{\gamma\chi\mathcal{C}} + 1 \right).
  \label{eq_linearly_incrementing_field}
\end{equation}

\begin{figure}
  \centering
  \includegraphics[width = 8.5cm]{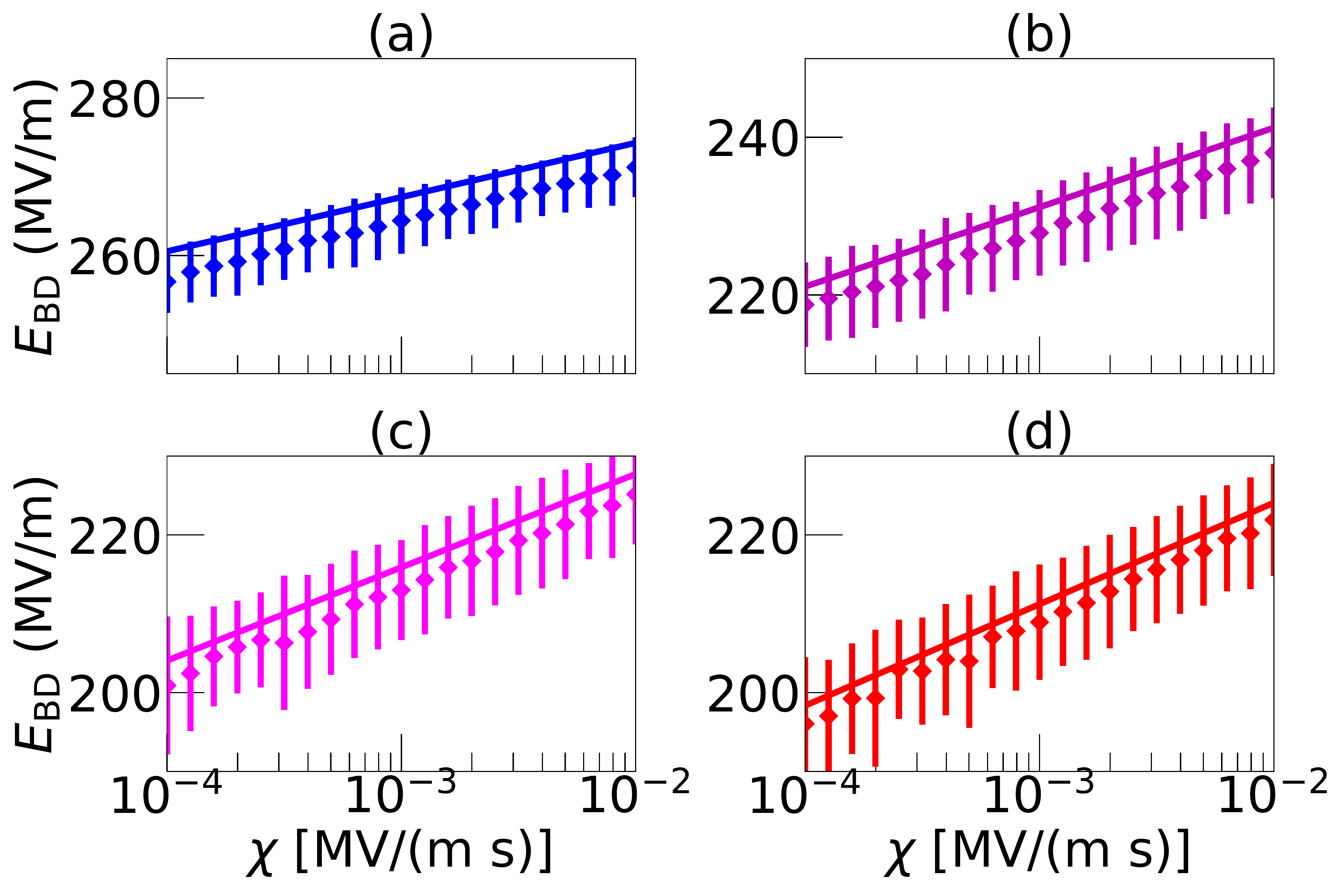}
  \caption
  {
    \label{fig_linearly_incrementing_field}
    Mean breakdown field as a function of the field increase rate
    for a linearly incrementing field,
    calculated using the metastable approximation
    [Eqs.~(\ref{eq_tau_metastable_final}) and
    (\ref{eq_linearly_incrementing_field}), solid line]
    and the simulation (squares with error bars),
    for temperatures of (a) 100, (b) 200, (c) 300, and (d) 400 K.
  }
\end{figure}

Figure \ref{fig_linearly_incrementing_field} shows $E_\text{BD}$ as a function
of the field increase rate $\chi$ for four temperatures.
For each temperature, $\mathcal{C}$ and $\gamma$ were found from a linear fit
to the results of the model for a constant field,
and were then used in Eq.~(\ref{eq_linearly_incrementing_field}).
Simulated breakdown fields are consistent with
(and, as expected, slightly lower than)
the results of Eq.~(\ref{eq_linearly_incrementing_field}).
All the mean breakdown times corresponding to data points in
Fig.~\ref{fig_linearly_incrementing_field} are greater than $2\times 10^5$
seconds of total field exposure time,
equivalent to $10^{12}$ typical 200 ns pulses.

\begin{figure}
  \centering
  \includegraphics[width = 8.5cm]{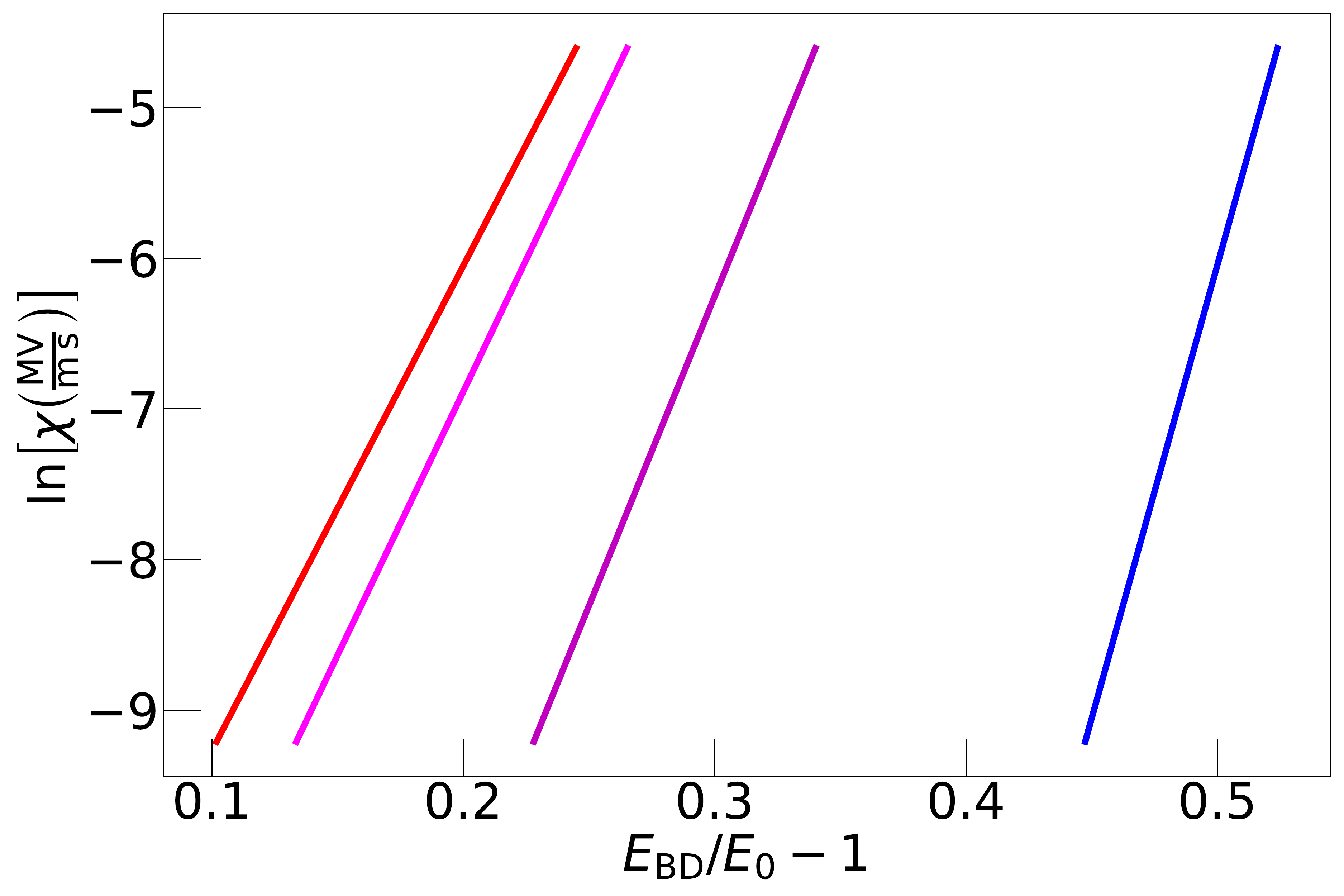}
  \caption
  {
    \label{fig_kissinger}
    Logarithm of the field increase rate as a function of
    $E_{\text{BD}}/E_0 - 1$,
    for temperatures of (lines from right to left) 100, 200, 300, and 400 K.
    Here $E_{\text{BD}}$ is the mean breakdown field,
    and $E_0$ is a reference field of 180 MV/m
    [as in Eq.~(\ref{eq_linear_ln_tau})].
    The slopes of the lines, from right to left, are 62.3, 41.7, 35.4, and 32.6,
    which are the values of $\gamma$ in Eq.~(\ref{eq_linear_ln_tau})
    for each of the corresponding temperatures.
  }
\end{figure}

It can be shown from Eq.~(\ref{eq_linearly_incrementing_field}) that $\ln\chi$
is a linear function of $E_\text{BD} / E_0 - 1$,
and that the slope of this function is $\gamma$,
as shown in Fig.~\ref{fig_kissinger}.
Hence, the value of $\gamma$, for a given structure at a given temperature,
can be found experimentally by measuring $E_\text{BD}$.
BDRs in the MDDF model are analogous to chemical reaction rates in
singly-activated kinetic scenarios.
The driving force for the transition is provided in the MDDF model by $E$,
instead of the temperature $T$ in the chemical reactions,
with $\gamma$ defining the sensitivity of the BDR to the electric field,
in the same way that the activation energy defines the sensitivity of the
reaction rate to the temperature.
The theoretical results described in Fig.~\ref{fig_kissinger} can serve as a
basis for a future experiment,
which will assist in identifying the controlling kinetics.
The proposed experiment is analogous to the Kissinger method,\cite{blaine12}
where the activation energy of a chemical reaction is found by increasing the
temperature of a specimen at several constant heating rates,
and measuring the exothermic peak temperature $T_m$
as a function of the heating rate.\cite{blaine12}

\section{Discussion and Conclusions} \label{sec_discussion}

The mobile dislocation density fluctuations (MDDF) model describes the plastic
response to an applied field,
via a stochastic process.
This early stage evolution can nucleate consequent dynamics,
which are described by other
models.\cite{anders08, pohjonen11, nordlund12, anders14, zadin14, vigonski15}
In addition,
the model defines some unique features of breakdown nucleation,
which have not been directly treated by previous models.
First,
breakdown is a critical process,
which develops within several tens of nanoseconds
for parameter values around those of the nominal parameters,
see Section \ref{sec_proposed}.
This can explain why pre-breakdown surface modifications are not observed
in samples,
regardless of proximity to the breakdown sites and time of exposure
to the field.
Secondly,
breakdown occurs deterministically for electric fields greater than $E_b$,
at which a bifurcation occurs where $\rho_*$ and $\rho_c$ merge.
Finally,
when the time to nucleate breakdown is comparable to the pulse length,
the BDR exhibits both a Poissonian and a non-Poissonian regime
within each pulse, see Fig.~\ref{fig_bd_distribution_within_pulse}.
For the nominal parameter set, this time is on the order of several tens of
nanoseconds,
suggesting BDR reduction for pulse lengths that are shorter than $O$(10 ns).

In addition to qualitative observations,
the MDDF model yields quantitative estimates,
following calibration of the unkown physical parameters,
which are in agreement with experimental results.
The agreement of the stochastic analysis and simulation results enable the use
of the former in cases where running the simulation is prohibitively long.
This expands the range of parameters and scenarios in which the predictions
of the MDDF model can be applied and put to test.
The model was used to predict BDRs outside the currently available experimental
data,
see Fig.~\ref{fig_linear_vs_square} and Section \ref{sec_proposed}.
Experiments conducted over these ranges,
where the predictions of the MDDF model and previous models diverge,
can serve to distinguish between models.

\begin{figure}
  \centering
  \includegraphics[width=8.5cm]{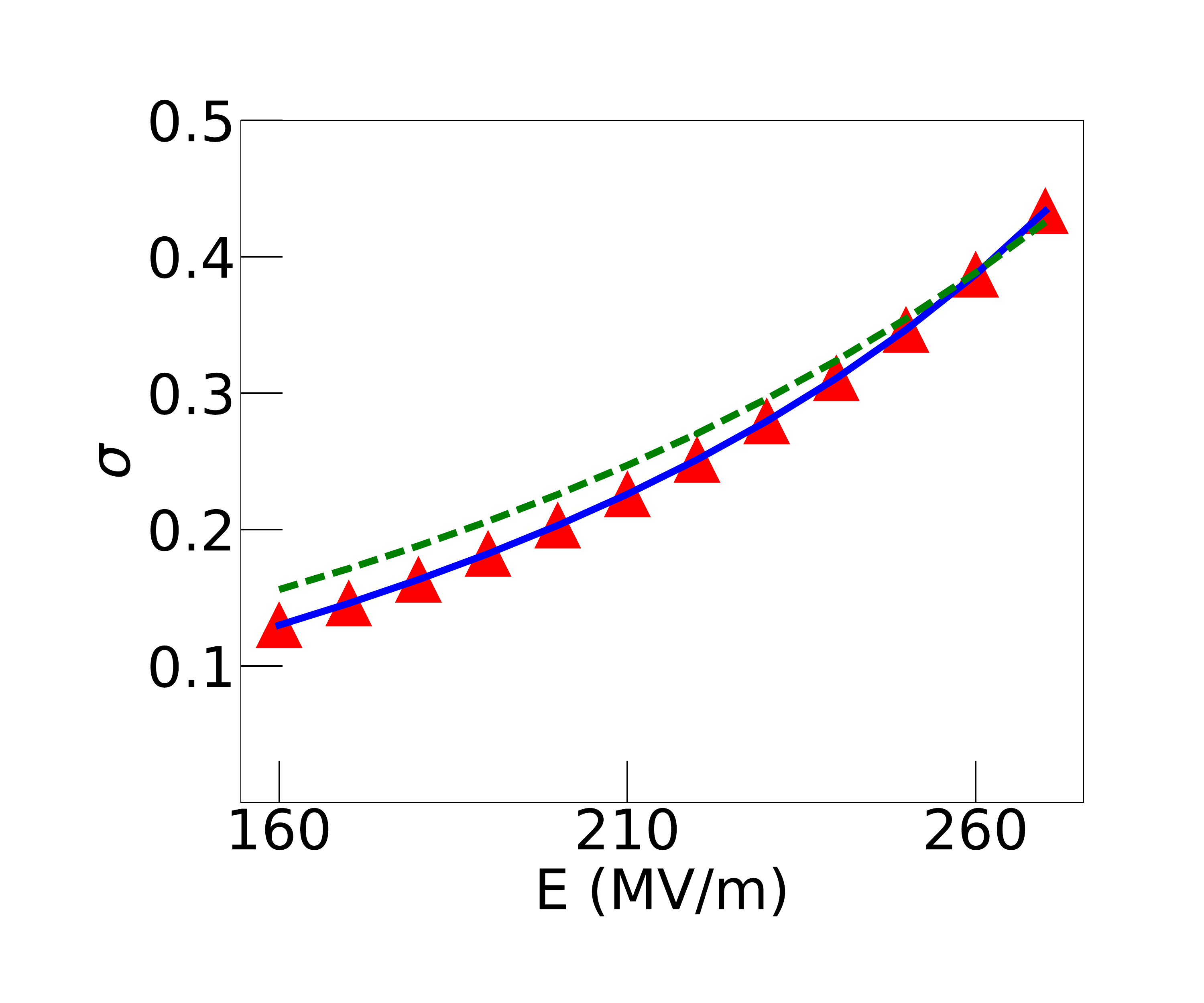}
  \caption
  {
    \label{fig_std}
    Standard deviation of the QSD as function of the electric field,
    calculated numerically from the QSD of the metastable approximation
    [Eq.~(\ref{eq_qsd}), solid line],
    calculated by treating the QSD approximately as a Poisson distribution
    [Eq.~(\ref{eq_poisson_mean}), dashed line],
    and found from the simulation (triangles).
  }
\end{figure}

Apart from predicting BDRs,
the MDDF model can be used to characterize aspects of a system
prior to breakdown.
Specifically,
the QSD of the mobile dislocation density (see Section \ref{sec_stochastic})
can be calculated with the metastable approximation,
or from numerical simulations.
The standard deviation of the QSD
is found to be an increasing function of the electric field,
and can be calculated directly from the QSD for each field,
see Fig.~\ref{fig_std}.
In addition,
a simpler expression of the standard deviation is derived by assuming that the
QSD can be approximated as a Poisson distribution in the vicinity of $n_*$
(see Section \ref{sec_stochastic}),
and therefore it can be estimated as the square root
of $\mathcal{R}$ in  Eq.~(\ref{eq_poisson_mean}).
This increase with field may be observed experimentally,
by measuring acoustic emission signals,
or by measuring the dark current between the cathode and anode
as a function of the applied electric field.\cite{weiss07}
This would allow the development of methods to detect early warning signals of
imminent breakdowns.\cite{scheffer09}

Such methods can be useful as part of the conditioning or other operational
schemes.
At present,
conditioning involves both extrinsic processes in which contaminants
are removed from the surface,
and intrinsic processes in which the surface structure
of the cathode metal is modified.\cite{wuensch13}
For example, in the CLIC project,
conditioning is done in a test stand reproducing the characteristics
of the application setup,
and typically takes six months to complete.\cite{degiovanni16}
Combining the ability to monitor early warning signals,
together with an understanding of the physical mechanism underlying
conditioning,
may allow the design of an improved conditioning procedure.

To conclude,
a theoretical link between fluctuations in the mobile dislocation density,
and its stochastic response to an external field,
is offered as a source for the critical process of breakdown under extreme
electric fields.
The MDDF model developed from this theory is analyzed and shown to provide a good
fit to a wide set of experimental data,
most of which was made available through the CLIC collaboration,
and to direct microscopic observations characterizing the dislocation structure
in electrodes.
Using the model,
expected responses in performed and planned experimental scenarios are
presented.
We suggest that experiments, utilizing temperature and drive rate variations,
can lead to significant improvement in the ability to identify specific
mechanisms controlling the critical transition which leads to eventual
breakdown.
In addition,
estimates are made of pre-breakdown changes in the evolution of dislocations.
Such changes may lead to an observable pre-breakdown signal,
which is currently under investigation.

\begin{acknowledgments}
  \emph{Acknowledgments.}
  We acknowledge K. Nordlund, F. Djurabekova, W. Wuensch, S. Calatroni,
  and J. Paszkiewicz for helpful discussions and providing data for
  Fig.~\ref{fig_fits}.
  Samples for Figs.~\ref{fig_sessile_dislocations} and \ref{fig_cells}
  were provided through the CLIC collaboration,
  with assistance from W. Wuensch, E. Rodriguez Castro, and I. Profatilova.
  We acknowledge funding from the PAZI foundation.
\end{acknowledgments}

\appendix

\section{Kinetic Monte-Carlo simulation} \label{sec_simulation}

In this appendix we describe the Kinetic Monte-Carlo simulations used to
describe the time evolution of the system and to compute the QSD and $\tau$.
The simulations implement a Gillespie algorithm,
tracking a single-step biased random walker along the $n$ axis.
The time spent between adjacent steps is randomly selected
from the exponential distribution,
$P = e^{-t/T}/T$,
with $T = (\lambda_n + \mu_n)^{-1}$ the average time spent between steps
in the state $n$.

\begin{figure}
  \centering
  \includegraphics[width=8.5cm]{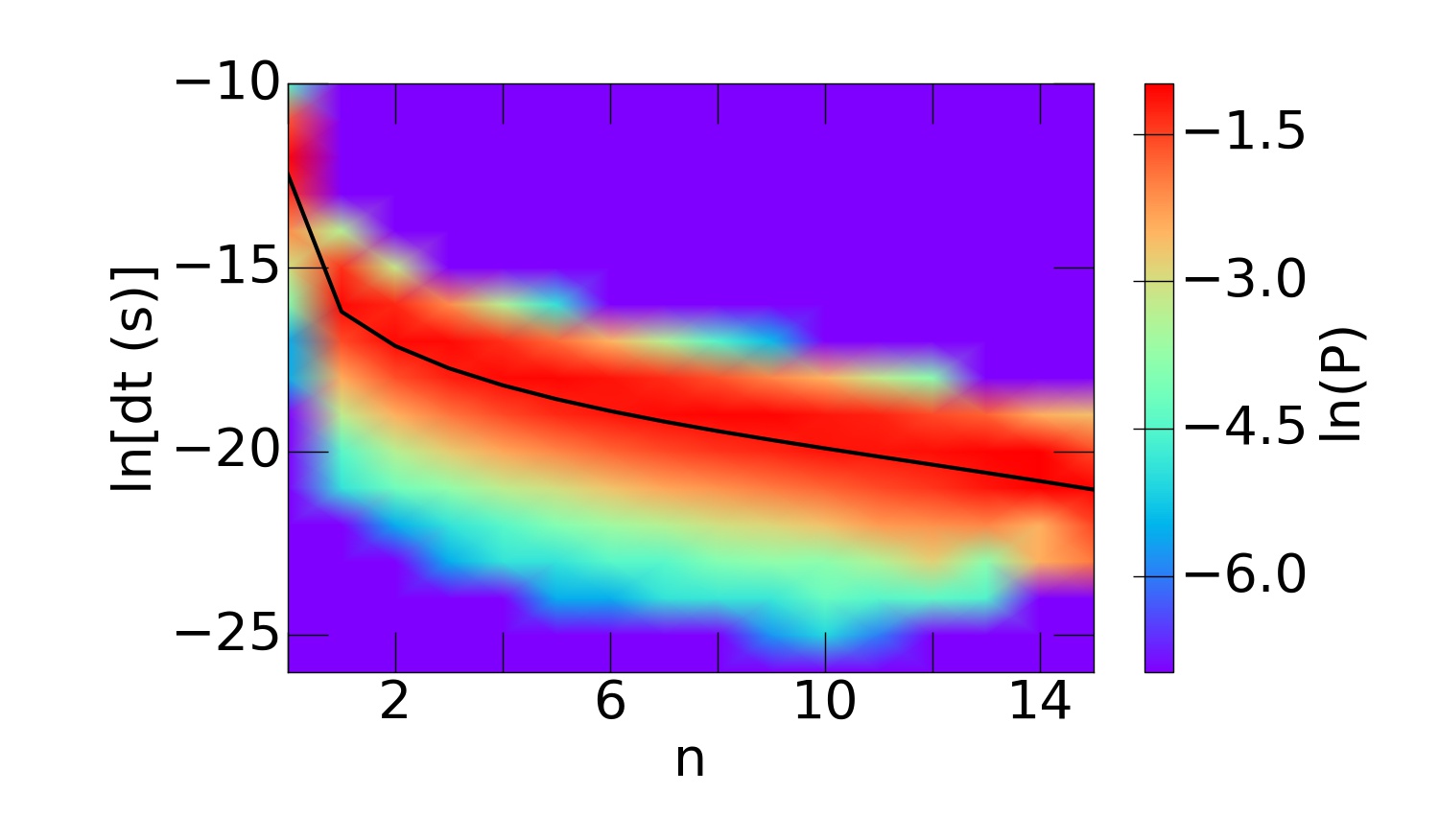}
  \caption
  {
    Probability distribution of the amount of time spent every step,
    for every $n$.
  }
  \label{fig_count_hists}
\end{figure}

To verify the time calculations in the simulations,
a histogram was created detailing, for every value of $n$,
the distribution of the amount of time spent in every step
in which the system was in state $n$.
This is shown in Fig.~\ref{fig_count_hists}.
The line in the figure represents the theoretical mean time spent at every step.
We find that the time distribution is as expected.

To find $\tau$ as a function of the electric field,
the simulations were run at least $10^2$ times for a number of electric fields,
with a reflecting state at $n = 0$
(i.e., $\mu_0 = 0$)
and an absorbing state at $n_c$,
as was assumed in the exact and metastable analyses.
The time spent at each step,
from the beginning of the run until reaching $n = n_c$,
was summed in order to find the breakdown time of each run.
The mean breakdown time $\tau$ was then calculated by averaging
the results of all the runs for that electric field,
and the error in $\tau$ was calculated as the standard deviation of the results.
In order to find the QSD,
the simulations were run, beginning at $n = 0$,
for at least $10^{11}$ steps every time, for a number of electric fields.

\section{LSQ Fit} \label{sec_fit}

The logarithm of the BDR in the reference data set depends linearly on
the magnitude of the electric field [see Fig.~\ref{fig_fits}(c)].
Comparing this relation to Eq.~(\ref{eq_linear_ln_tau})
yields $\gamma$ = 31.4 (at $T$ = 300 K).
Using the converstion ratio given by Eq.~(\ref{eq_conversion_ratio}),
we also find that $\log_{10}[R_\text{hot}\,(\text{bpp}/\text{m})] = -5.89$,
where $R_\text{hot}$ is the BDR at $E$ = 180 MV/m and $T$ = 300 K.

To incorporate the temperature dependence of the BDR into the parameter fit,
we note that the leftmost data set in Fig.~\ref{fig_fits}(b)
and the single measurement at 45 K must be fitted
with the same $\beta$ since they were both measured in the same structure,
or, equivalently, the electric field must be scaled identically in both sets.
Scaling the fields,
so that the data of the leftmost set matches the fit of the reference set,
yields a scaled field of 300 MV/m for the measurement at 45 K.
Comparing the BDRs of this measurement and the reference set, we have
$\log_{10}(R_\text{hot}/R_\text{cold}) \approx 1.8$,
where $R_\text{cold}$ is the BDR at $E$ = 300 MV/m and $T$ = 45 K.

Defining $Q_1 = \log_{10}[R_\text{hot}\,(\text{bpp}/\text{m})]$ and
$Q_2 = \log_{10}(R_\text{hot}/R_\text{cold})$,
the LSQ fit was carried out by finding the lowest value
of the total quality measure $Q$ in the four-parameter phase space
($E_a$, $\Omega$, $\kappa$, $\beta$),
where
\begin{align}
  Q &= 4\left(\frac{\gamma}{31.4} - 1\right)^2
  + \left(\frac{Q_1}{-5.89} - 1\right)^2 \label{eq_lsq} \\
  &+ \left(\frac{Q_2}{1.8} - 1\right)^2
  + \left(\frac{n_{c,\text{hot}}}{30} - 1\right)^2 \nonumber \\
  &+ \left(\frac{n_{c,\text{cold}}}{30} - 1\right)^2
  + \left(\frac{E_a}{0.1\,\text{eV}} - 1\right)^2. \nonumber
\end{align}
Here, $n_{c, \text{hot}}$ is $n_c$ at $E$ = 180 MV/m and $T$ = 300 K,
and $n_{c, \text{cold}}$ is $n_c$ at $E$ = 300 MV/m and $T$ = 45 K.
These were included in the quality measure because the statistical
mean-field analysis and the metastable approximation are valid
only when $n_c \gg 1$.
The target values for these measures were chosen because they
are larger then the calculated values in the whole region of the
phase space where the parameters have plausible values,
so that the larger $n_{c, \text{hot}}$ and $n_{c, \text{cold}}$ are,
the smaller $Q$ is.
Similarly,
the lowest theoretical estimate of $E_a$ to date is $\sim$ 0.1 eV,\cite{zhu08}
while within the phase space where the other parameters have plausible values
it is found that $E_a < 0.1$.
Therefore, $E_a$ was added as a quality measure with a target value of 0.1 eV,
so that the greater $E_a$ is, the smaller $Q$ is.
The value of $\gamma$ was given greater weight than the other measures
as its relative error is smaller by $\sim 4$ than that of the other measures.

\begin{table}[!hb]
  \caption
  {
    \label{tab_params}
    Measures and target values for the LSQ fit of the physical parameters
    [Eq.~(\ref{eq_lsq})].
  }
  \begin{ruledtabular}
    \begin{tabular}{ccD{.}{.}{2.2}D{.}{.}{1.1}}
      Measure & Description & \multicolumn{1}{c}{Target value} &
      \multicolumn{1}{c}{Weight} \\
      \hline
      $\gamma$ & Fit to Eq.~(\ref{eq_linear_ln_tau}) & 31.4 & 4 \\
      $Q_1$ & $\log_{10}[R_\text{hot}\,(\text{bpp}/\text{m})]$ & -5.89 & 1 \\
      $Q_2$ & $\log_{10}(R_\text{hot}/R_\text{cold})$ & 1.8 & 1 \\
      $n_{c,\text{hot}}$ & $n_c(180\,\text{MV/m}, 300\,\text{K})$ & 30 & 1 \\
      $n_{c,\text{cold}}$ & $n_c(300\,\text{MV/m}, 45\,\text{K})$ & 15 & 1 \\
      $E_a$ & Activation energy & 0.1\,\text{eV} & 1
    \end{tabular}
  \end{ruledtabular}
\end{table}

Table \ref{tab_params} summarizes the measures of the LSQ fit,
their target value, and the relative weight of each measure.
The optimal set of parameters found from the fit is the nominal set
described in Section \ref{sec_mean}.

%\bibliography{main}

%merlin.mbs apsrev4-1.bst 2010-07-25 4.21a (PWD, AO, DPC) hacked
%Control: key (0)
%Control: author (8) initials jnrlst
%Control: editor formatted (1) identically to author
%Control: production of article title (-1) disabled
%Control: page (0) single
%Control: year (1) truncated
%Control: production of eprint (0) enabled
%

\end{document}